\documentclass[english,aps, pra, superscriptaddress,onecolumn,11pt,tightenlines,notitlepage]{revtex4-1}

\usepackage{etex}
\makeatletter
\def\alloc@#1#2#3#4#5%
 {\ifnum\count1#1<#4
    \allocationnumber\count1#1
    \global\advance\count1#1\@ne
    \global#3#5\allocationnumber
    \wlog{\string#5=\string#2\the\allocationnumber}%
  \else\ifnum#1<6
    \def\etex@dummy@definition{}
    \begingroup \escapechar\m@ne
    \expandafter\alloc@@\expandafter{\string#2}#5%
  \else\errmessage{No room for a new #2}\fi\fi
 }
\makeatother

\usepackage[T1]{fontenc}
\usepackage[utf8]{inputenc}
\usepackage{color}
\definecolor{note_fontcolor}{rgb}{0.800781, 0.800781, 0.800781}
\usepackage[english]{babel}
\usepackage{lmodern}
\usepackage{mathtools}
\usepackage{amsmath}
\usepackage{amssymb}
\usepackage{graphicx}
\usepackage{easybmat,multirow,bigdelim}
\usepackage{xcite}
 \usepackage{amsthm}
 \usepackage[unicode=true,pdfusetitle, bookmarks=true,bookmarksnumbered=false,bookmarksopen=false,pdfborder={0 0 1},backref=false,colorlinks=false]{hyperref}

 \newtheorem{theorem}{Theorem}
 \newtheorem{lemma}[theorem]{Lemma}
 \newtheorem{definition}{Definition}
\newtheorem{fact}[theorem]{Fact}
\newenvironment{proofof}[1]{\noindent{\bf Proof}
of #1:\hspace*{0.5em}}{$\square$\bigskip}

\newcommand{\ZZ}{\ensuremath{\mathcal{Z}}}
\newcommand{\RR}{\textrm{R}}
\newcommand{\lr}{\leftrightarrow}

\DeclareMathOperator{\R}{R}
\DeclareMathOperator{\Ima}{Im}

\newcommand{\ket}[1]{|#1\rangle}

\newcommand*\hexbrace[2]{%
\underset{#2}{\underbrace{\rule{#1}{0pt}}}}
  
\newcommand{\onote}[1]{\textcolor{blue}{ {\textbf{(Or:}
#1\textbf{) }}}}
\renewcommand{\onote}[1]{}
\begin{document}

\title{When a local Hamiltonian must be frustration-free}
 \author{O. Sattath}
 \affiliation{Computer Science Division, University of California, Berkeley, CA, 94720, USA }

 \author{S. C. Morampudi}
 \affiliation{Max-Planck-Institut f\"{u}r Physik komplexer Systeme, 01187 Dresden, Germany}

 \author{C. R. Laumann}
 \affiliation{Department of Physics, University of Washington, Seattle, WA, 98195, USA}

 \author{R. Moessner}
 \affiliation{Max-Planck-Institut f\"{u}r Physik komplexer Systeme, 01187 Dresden, Germany}



\begin{abstract}
{A broad range of quantum optimisation problems can be phrased as the question whether a specific system has a ground 
state at zero energy, i.e.\ whether its Hamiltonian is frustration free. 
Frustration-free Hamiltonians, in turn, play a central role for constructing and understanding new phases of matter in 
quantum many-body physics. 
Unfortunately, determining whether this is the case 
is known to be a complexity-theoretically intractable problem. This makes it highly desirable to search for 
efficient heuristics and algorithms in order to, at least, partially answer this question. 
Here we prove a general criterion -- a sufficient condition -- under which a local Hamiltonian 
is guaranteed to be frustration free by lifting Shearer's theorem from classical probability theory to the quantum world. Remarkably, evaluating this condition proceeds via a fully
classical analysis of a hard-core lattice gas at negative fugacity on the Hamiltonian's interaction graph which, as a statistical mechanics problem, is of interest in its own right.
We concretely apply this criterion to local Hamiltonians on various regular lattices, while bringing to bear the tools of spin glass physics which permit us to obtain new bounds on the SAT/UNSAT transition in random quantum satisfiability.
These also lead us to natural conjectures for when such bounds will be tight, as well as to a novel notion of universality for these computer science problems. 
Besides providing concrete algorithms leading to detailed and quantitative 
insights, this underscores the power of marrying classical statistical mechanics with quantum computation and complexity theory.}
\end{abstract}
 \maketitle

\begin{figure}[htbp]
\includegraphics[width=276pt]{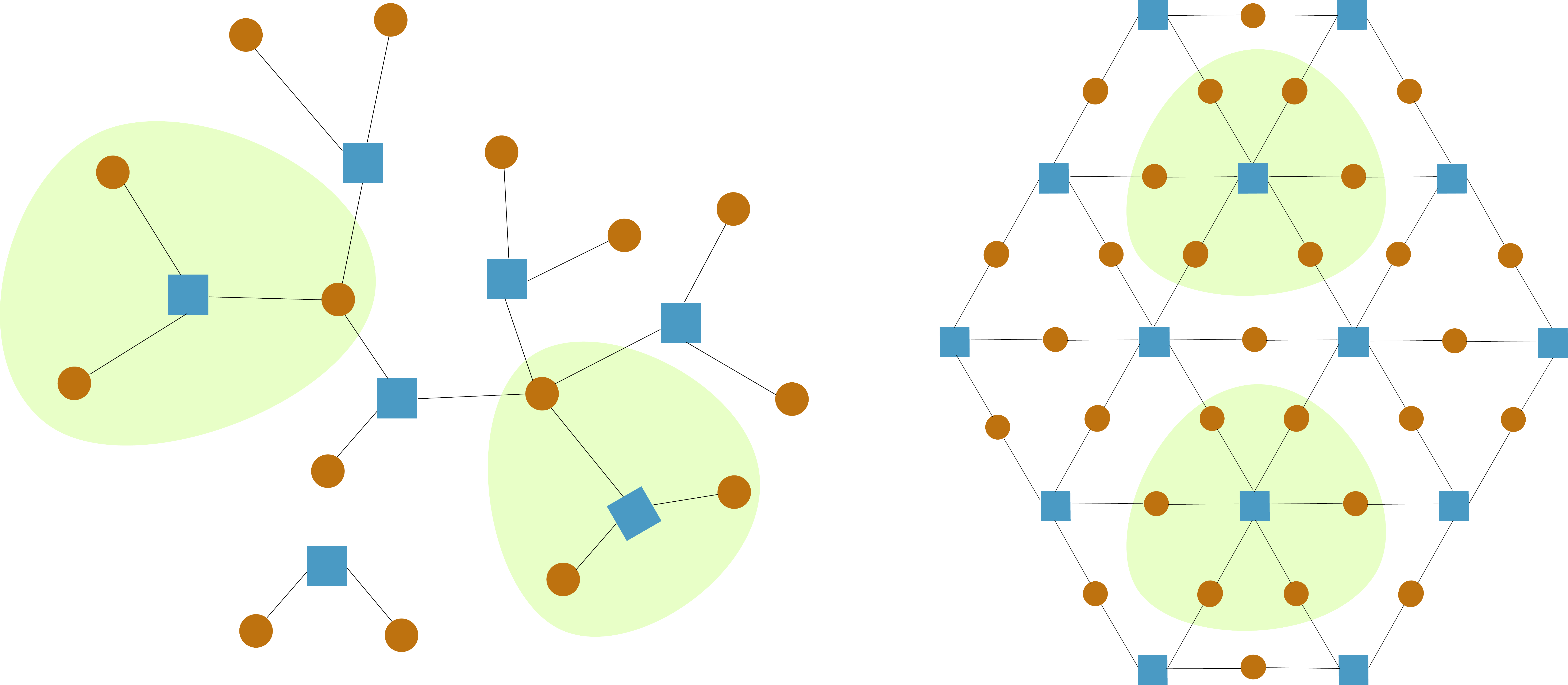}
\protect\caption{\label{fig:intgraph}
Left: The interaction graph of a $k=3$-local Hamiltonian. The degrees of freedom, qudits, are denoted by circles; the squares, which indicate the operators $\Pi_i$ appearing in $H$, are joined to the qudits on which they act. 
In the hard-core lattice gas mapping, each square may be occupied by a particle covering also the adjacent circles (shaded blobs), on which it must not overlap with another particle.
Right: A $k=6$ interaction graph forming a triangular lattice of the operators $\Pi_i$. The corresponding hard-core lattice gas corresponds to the hard hexagon model, which is exactly solvable.}
\end{figure}

A%
n overwhelming majority of systems of physical interest can be described via local Hamiltonians:
\begin{equation}
H = \sum_{i=1}^M \Pi_i
\label{eq:localham}
\end{equation}
Here, the `$k$-local' operator $\Pi_i$ acts on a $k$-tuple of the microscopic degrees of freedom, best
thought of as qudits for the computer scientists among our readers, or spins for the physicists.
The $M$ operators define an  interaction (hyper-)graph $G$, as displayed in Fig.~\ref{fig:intgraph}.

A surprisingly diverse and important class of such model Hamiltonians is defined by the additional property of being \emph{frustration-free}: the ground state $\ket{\psi}$ of $H$ is a simultaneous ground state of each and every $\Pi_i$. 
This class comprises both commuting Hamiltonians -- for which $[\Pi_i,\Pi_j]=0\ \  \forall i,j$ -- such as the toric code, general quantum error correcting codes and Levin-Wen models\cite{Kitaev2003,Gottesman2010,LevinWen}; as well as non-commuting ones, such as the AKLT and Rokhsar-Kivelson models\cite{AKLT1987,RKOriginal,CastelnovoRK}. 
Their particular usefulness is also related to the fact that 
many of these examples can be viewed as `local parent Hamiltonians' for generalized matrix product states \cite{perez2008peps}.
In general,  frustration-free conditions provide analytic control of ground state properties in otherwise largely inaccessible quantum problems. 

Determining whether a given Hamiltonian $H$ is frustration-free is well known in quantum complexity 
theory as the quantum satisfiability (QSAT) problem. QSAT is widely believed to be intractable, in the sense that
no general purpose classical or quantum algorithm can efficiently determine whether a 
given Hamiltonian is frustration-free (`satisfiable'). The technical statement is that QSAT is QMA$_{1}$-complete~\cite{bravyi2011efficient}, even when restricted to qubits and $k=3$~\cite{GossetN13}, or when the interactions is between neighboring qudits on a line~\cite{aharonov_power_2007}.

Fortunately, such hardness results only apply in the worst case. For instance, in contrast to the hardness result, it is immediately obvious that a fully disconnected interaction graph $G$ can 
be analysed efficiently by considering each term $\Pi_i$ individually.

The central result reported here is a sufficient combinatorial criterion for a local Hamiltonian $H$ to be frustration free. 
In fact, we provide a lower bound for the dimension of the satisfying subspace. This amounts to a
generalisation of Shearer's theorem\cite{Shearer1985} from classical probability theory to the quantum world.

We first formulate the result in Theorem 1, followed by an intuitive explanation of its content, with
a technical proof relegated to the supplemental information. 
We then turn to applying Theorem 1, for which
we enlist the results available on statistical mechanics of hardcore objects with negative fugacity on the interaction graph
to deduce statements regarding QSAT, producing new bounds on the satisfiability threshold for a large 
class of one, two and three dimensional interaction graphs. With help of the cavity method, we are able to conjecture improved  lower bounds for the satisfiability of QSAT on regular and Erd\H{o}s-R\'enyi random graphs, 
canonical models for quantum constraint optimization problems.
These statement hold just as well for classical satisfiability -- but for some of these classical models better bounds are known~\cite{achlioptas03threshold,rathi10bounds}.  

We close with an outlook, including a discussion of the role of a universality which emerges in our analysis, as well as conjectures on when our results are exact or tight. 

\begin{theorem}
	Given a local projector Hamiltonian $H$ as in Eq.~\eqref{eq:localham} with interaction graph $G$ and relative projector rank $p = \R(\Ima \Pi_i)$ for all $i$, then 
	\begin{align*}
	\R(\ker H) \ge \ZZ(G,-p) > 0
	\end{align*}
	if $\ZZ(G,-p')> 0$ for all $0 \le p' \le p$.
\label{thm:qshearer_qsat}
\end{theorem}
Without loss of generality, we have taken the terms $\Pi_i$ to be projectors so that the 
satisfiability condition reduces to the presence of a zero energy ground state \footnote{This 
may always be done by shifting and deforming the eigenvalues of the local Hamiltonian terms 
without influencing the frustration free ground state space.}.

Here, the relative dimension of a subspace $X$ of the full Hilbert space $\mathcal{H}$ is 
simply $\R(X)=\frac{\dim X}{\dim \mathcal{H}}$, and, $\ZZ(G,\lambda)$ is the partition function 
for a hardcore lattice gas of fat particles living on the (hyper)-edges of the interaction graph at 
fugacity $\lambda$ (see Fig.~\ref{fig:intgraph}). More precisely,
\begin{align}
\label{eq:Zhardcore}
	\ZZ(G,\lambda) = \sum_{\{n_i\}} \lambda^{\sum n_i} \prod_{i \lr j} (1 - n_i n_j)
\end{align}
where $i \lr j$ runs over all projectors which share qudits and the sum runs over 
occupations $n_i = 0,1$ of the lattice gas.

\paragraph{Intuition---}
\label{sec:Quantum-Shearer-bound}

Let us first consider the case of classical projectors, $\Pi_i$, diagonal in the
standard computational basis, where we can provide a simple pictorial representation of the content of the Shearer bound. The operator $\prod_i(1-\Pi_i)$ projects onto the zero-energy space of the Hamiltonian. If it is non-zero the Hamiltonian is frustration-free. Moreover, its expectation value in the maximally mixed (infinite temperature) state gives the relative dimension of the zero-energy space,
\begin{align}
\label{eq:reldimclassical}
    R(\ker H) &= \overline{\prod_{i=1}^M (1-\Pi_i)}
\end{align}
If all of the projectors act on different qudits, then, this expectation value simply factors,
\begin{align*}
R(\ker H) &= \prod_{i=1}^M \overline{(1-\Pi_i)} = (1-p)^M
\end{align*}
where $p = \overline{\Pi_i}$ is the relative dimension of $\Pi_i$. 
This is non-zero so long as $p < p_c = 1$.

\begin{figure}[htbp]
\includegraphics[width=276pt]{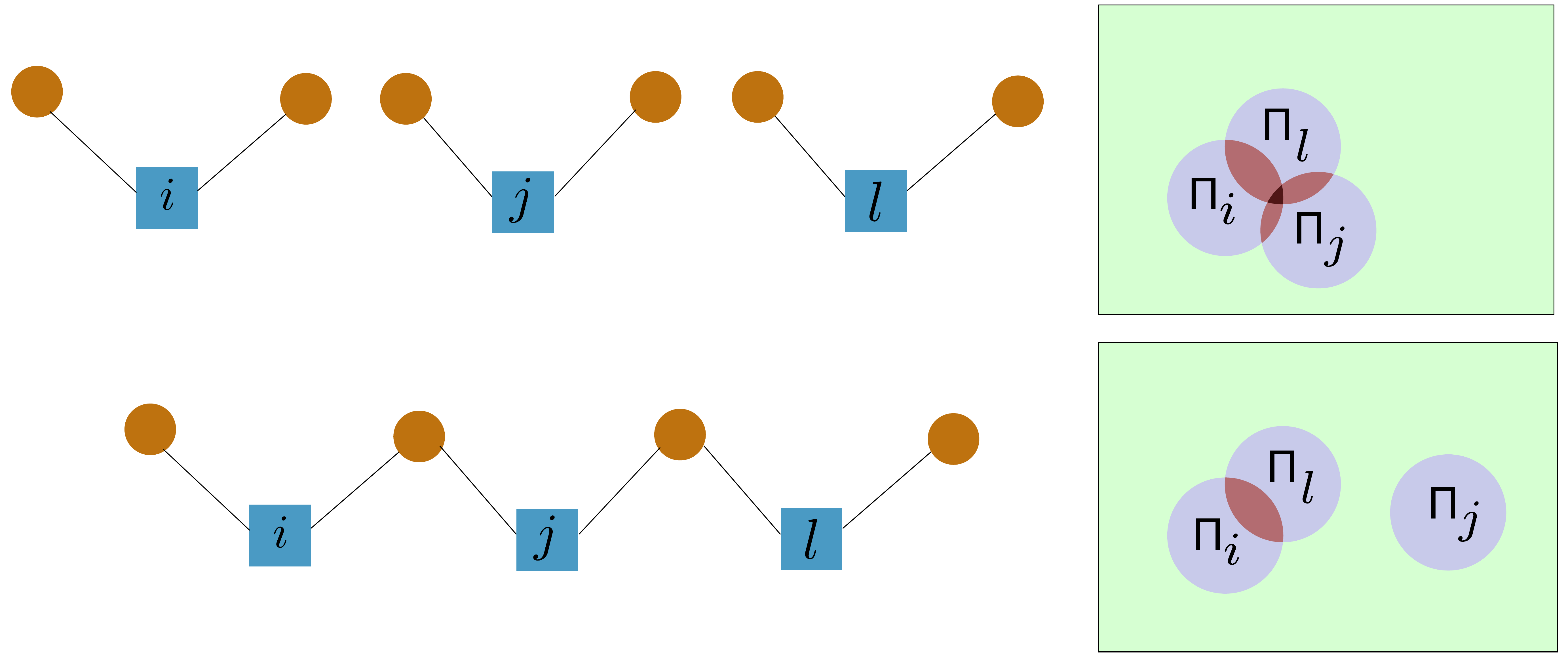}
\protect\caption{\label{fig:shearBubb}
Shearer bound for dependent projectors.
Projectors are independent if they do not share any qudit (top left), and there will always exist configurations 
violating any combination of such projectors simultaneously. In the Venn diagram (top right), this is represented by the mutual intersections of the shaded circles, each of which denotes the fraction of configuration space penalised by the projector it is labeled with. By contrast, for dependent projectors (bottom left), such an overlap is not guaranteed.
A lower bound for the relative dimension of the satisfying space can be obtained by assuming the projectors  $\Pi_{i}$ and $\Pi_{l}$ do not share any violating configurations with the projector $\Pi_{j}$ with which they share a qudit. (bottom right)}
\end{figure}

Now suppose that some of the projectors share qudits; we call such pairs of projectors dependent, as they introduce dependence into the expectation value in Eq.~\eqref{eq:reldimclassical}. 
When can we, nevertheless, guarantee that $R(\ker H)$ remains positive? 
The essence of the Shearer bound is captured by the Venn diagram in Fig.~\ref{fig:shearBubb}, where each bubble represents the fraction of configuration space `knocked out' by a projector, with overlapping bubbles representing configurations multiply penalised by the corresponding projectors. 

The area covered by the bubbles is largest if they do not overlap, corresponding to a large number of violating configurations; 
we thus expect a lower bound on $R(\ker H)$ when they are fully disjoint. 
To calculate this lower bound, we expand the product in Eq.~\eqref{eq:reldimclassical} and introduce a collection of occupation variables $n_i = 0,1$ which indicate the presence of $\Pi_i$ in each term of the expansion,
\begin{align}
	\label{eq:inclusionexclusion}
	R(\ker H) &= \sum_{\{n_i\}} (-1)^{\sum n_i} \overline{\Pi_1^{n_1} \cdots \Pi_M^{n_M}}
\end{align}
If two dependent projectors, $\Pi_i$ and $\Pi_j$, are thus occupied, then that term is zero when their bubbles are made disjoint. Otherwise, it is given by $p^{\sum n_i}$. Thus,
\begin{align}
\label{eq:allgoodineq}
	R(\ker H) &\ge \sum_{\{n_i\}} (-p)^{\sum n_i} \prod_{i \leftrightarrow j}(1-n_i n_j) \nonumber\\
	&= \ZZ(G,-p)
\end{align}
where $i \leftrightarrow j$ runs over projectors that share qudits. 

We have derived inequality \eqref{eq:allgoodineq} under the assumption that it is possible to make the dependent bubbles disjoint. 
If $p$ is small enough, this is always the case; in the Venn diagram, the bubbles can be made disjoint without covering more area than the total space contains. 
Shearer~\cite{Shearer1985,Scott2005} showed that the above intuitive lower bound is correct for classical projectors so long as $p \leq p_c$ where $p_c$ is the first zero of $\ZZ(G,-p)$. 
This is the classical analog of our quantum generalization, Theorem 1.

The classical sketch above makes little sense for non-commuting quantum projectors. In the language of probability, this reflects the failure of the inclusion-exclusion principle (Eq.~\eqref{eq:inclusionexclusion}) for the relative dimension of vector spaces. 
Nonetheless, the result holds; the proof -- our fundamental technical advance -- is provided in Appendix~\ref{sec:shearer_proof}. 

\paragraph{Statistical mechanical transcription---}
A remarkable aspect of Theorem 1 is that it maps the \emph{quantum} satisfiability problem onto the \emph{classical} statistical mechanical problem of determining the position of the first negative fugacity zero $\lambda_c(G)$ of the partition function $\ZZ(G,\lambda)$. 
In general, evaluating $\ZZ$ is computationally hard
\cite{[{Technically, computing the independent set polynomial 
 is $\sharp$P-hard almost everywhere.}]Hoffmann2010}.
In the thermodynamic limit (number of qudits $N\to\infty$), the first zero $\lambda_c(G)$ can be identified with a well known critical point $\lambda_c(G_\infty)$ of the hard-core lattice gas referred to as the \emph{hard-core singularity}. 
The critical fugacity $\lambda_c(G_\infty)$ upper bounds the $\lambda_c$ of any finite subgraph by monotonicity \cite{Shearer1985}. 
Thus, Theorem 1 implies that all subgraphs of $G_\infty$ are frustration free so long as the relative rank of the projectors satisfies $p < - \lambda_c(G_\infty)$.

The hard-core singularity has been studied extensively in the statistical mechanical literature \cite{Groeneveld, Gaunt1965, Gaunt1967, heilmann1972, Baxter1980, Poland1984, PhysRevA.36.760, todo1999transfer, Lai1995, PhysRevE.60.6323} and its location is known for many infinite lattices. 
Table~\ref{tab:critical_points} summarizes some of these critical values and indicates their translation into QSAT interaction graphs. 
In addition, there exist rigorous bounds on $\lambda_{c}$ which may be efficiently calculated on any sufficiently simple lattice \cite{Groeneveld}.
\begin{table}[h] \centering 
\caption{Summary of the critical threshold for various infinite interaction graphs} 
\begin{tabular}{ l l l l l l  } 
\hline 
Lattice & $p_c$ & Qudits & Projectors & $k$ \\
\hline \hline
1-D chain & $\frac{1}{4}$ ~\cite{movassagh2010unfrustrated} & vertices & edges & 2 \\
 
Triangular & $\frac{5 \sqrt 5-11}{2}$ ~\cite{Gaunt1967,Baxter1980,todo1999transfer} & edges & vertices & 6  \\

Square &  $0.1193$ ~\cite{Gaunt1965,todo1999transfer} & edges & vertices & 4\\

Square & $0.0889$~\cite{todo1999transfer}& vertices & edges & 2 \\

Checkerboard & $0.0688$~\cite{todo1999transfer} & black & red &  4 \\

Hexagonal  &  $0.1547$~\cite{todo1999transfer} & edges & vertices & 3 \\

Simple Cubic & $0.0744$~\cite{Gaunt1967} & edges & vertices & 6 \\

$t$-regular tree & $\frac{1}{4(t-1)}$~\cite{heilmann1972,coudron2012unfrustration} & vertices & edges & 2 \\
\hline
 \end{tabular}
 \label{tab:critical_points}
\end{table}
The singularity exhibits universal features just like standard phase transitions.
In particular, the free energy density $f(\lambda) = \frac{1}{N} \log \ZZ(\lambda)$ near the critical point $\lambda_c$ has universal exponents \cite{Poland1984, PhysRevA.36.760} due to its connection with the so-called \emph{Yang-Lee edge singularity} \cite{PhysRev.87.404, PhysRev.87.410, Shapir1982, Lai1995, PhysRevE.60.6323}.
Technically, this means that the leading non-analytic part of the free energy $f \sim \left(\lambda-\lambda_{c}\right)^{\phi_D}$ near $\lambda_c$ where $\phi_D$ is a non-integer critical exponent which depends only on the spatial dimension $D$ of the lattice \cite{Poland1984,PhysRevA.36.760,Lai1995}.
It is known analytically that $\phi_1 = 1/2$, $\phi_2 = 5/6$ \cite{Baxter1980,PhysRevLett.40.1610,PhysRevLett.54.1354} and $\phi_D = 3/2$ for $D$ above the upper critical dimension $D_C = 6$ \cite{Lai1995}.

The critical exponents control the lower bounds on the dimension of the ground space of Hamiltonians close to the critical threshold. 
Indeed, by Theorem 1, $R(\ker H) \ge e^{n f(\lambda)} \propto e^{a n (\lambda - \lambda_c)^{\phi_D} + O(\lambda - \lambda_c)}$ for $\lambda > \lambda_c$, so the critical exponents show up directly -- and perhaps somewhat unexpectedly -- as anomalous growth laws in $D=1,2$.
In the following, observation of the expected exponent in the random $k$-QSAT ensemble helps confirm the validity of the non-rigorous cavity analysis.

\paragraph{Examples---}
As a first application of Theorem 1, we calculate the critical threshold $\lambda_c$ for several infinite graphs. We do this using the cavity method, a well-known technique for studying statistical\cite{Mezard2001} and quantum \cite{laumannqcav,poulinqcav,hastingsqcav} models on infinite graphs which are locally tree-like. 
On trees and chains, the results are rigorously exact while on infinite random graphs they are less rigorous, but often just as exact.

The heart of the cavity method is to introduce an auxiliary system of messages which propagate in both directions between the hyperedges and the nodes of the interaction graph $G$. 
For the model given by Eq.~\eqref{eq:Zhardcore}, the derivation is straightforward, and quite intuitive at positive fugacity where the cavity messages correspond to marginal distributions for the gas particles - see Appendix~\ref{sec:cavity_derivations}. 
The resulting equations continue smoothly to the negative fugacity regime.
Adopting the convention that indices $i,j,...$ label hyperedges and $a,b,...$ label sites, the belief propagation (BP) equations%
\cite{[{For a recent pedagogical treatment, see Ch.~14 of }]Mezard:2009:IPC:1592967}
\begin{align*}
q_{a\rightarrow i} & =\dfrac{\lambda}{{\displaystyle \lambda+1+\sum_{j\epsilon\partial a\backslash i}\dfrac{l_{j\rightarrow a}}{1-l_{j\rightarrow a}}}} \nonumber\\
l_{i\rightarrow a} & =\dfrac{\lambda}{\lambda+{\displaystyle \prod_{b\epsilon\partial i\backslash a}\left(\dfrac{\lambda}{q_{b\rightarrow i}}-\lambda\right)}}
\end{align*}
where $l_{i\rightarrow a}$ and $q_{a \rightarrow i}$ give the probability that hyperedge $i$ is occupied in the absence of the connection to $a$ and in the absence of everything except $a$ respectively. 
We note that a special case of these equations has been used previously to study the number of dimer coverings of random graphs ($k=2$, positive fugacity) \cite{lenkamatchings}.

Given a solution of the BP equations, the cavity estimate of the free energy $F = \log \ZZ$ of the system is given by a sum of the free energies associated to the addition of individual elements
of the interaction graph, i.e.,
\begin{align}
F & ={\displaystyle \sum_{a}F_{a}+\sum_{i}F_{i}-\sum_{(ai)}F_{ai}}
\label{eq:free_energies}
\end{align}
where $F_{a}$, $F_{i}$ and $F_{ai}$ correspond to the change in
free energy due to the addition of sites $a$, hyperedges $i$ and the links $ai$ between them, 
respectively.
The full expressions may be found in Appendix~\ref{sec:cavity_derivations}.

\paragraph{The chain---}
On the infinite chain, the BP equations can be solved by uniform messages $q_{a\to i} = q$, $l_{i \to a} = l$. 
After some algebra and taking the root of the BP equations corresponding to $q(\lambda = 0) = 0$, one finds $l = q = 1 + \frac{1 - \sqrt{1+4\lambda}}{2\lambda}$. 
This expression suggests that $\lambda_c = -1/4$ as $q$ becomes complex for $\lambda < \lambda_c$. 
Indeed, it is easy to check that the free energy density $f = F/N$ Eq.~\eqref{eq:free_energies} has the expansion near $\lambda_c$:
\begin{align*}
f &= -\log 2 + 2 (\lambda - \lambda_c)^{1/2}  - 2 (\lambda - \lambda_c) + \cdots
\end{align*}
The free energy goes complex for $\lambda < \lambda_c$, which reflects the accumulation of partition function zeros, and indicates the failure of the lower bound. 

The free energy density is precisely $- \log 2$ at the critical point. 
For qudits of local dimension $q$, this provides a lower bound on the dimension of the ground space $\dim(\ker H) = q^N R(\ker H) \ge \left(\frac{q}{2}\right)^N$. In fact, careful application of Theorem~\ref{thm:qshearer_qsat} for finite chains agrees with the exact result $\dim(\ker H) = \left(\frac{q}{2}\right)^N (N+1)$ for open chains derived using matrix-product techniques \cite{Laumann:2010p7275,movassagh2010unfrustrated}.
It is interesting to note that the entropy per site is zero for qubits at criticality, but has a finite value for larger $q$.


\paragraph{Regular trees---}
The BP equations on infinite regular trees with site degree $t$ and hyperedge degree $k$ can be solved under the ansatz that all messages are the same. For algebraic details, see Appendix~\ref{sec:cavity_derivations}. 
The resulting critical threshold is
\begin{align}
\label{eq:lambdacreg}
\lambda_c = -\frac{1}{t-1} \frac{(k-1)^{k-1}}{k^k}
\end{align}
For $k=2$-local trees, this agrees with previous results obtained by matrix product states on trees\cite{coudron2012unfrustration}. 
\nopagebreak
As far as we are aware, Eq.~\eqref{eq:lambdacreg} provides a strictly better lower bound on satisfiability of infinite regular trees than any previous literature.
The corresponding expansion of the free energy density is
\begin{align}
\label{eq:freeenergytree}
f = f_0(t,k) + A(t,k) (\lambda - \lambda_c) + B(t,k) (\lambda - \lambda_c)^{3/2} + \cdots 
\end{align}
We thus recover the expected non-analyticity ($\phi=3/2$) in the free energy density for the infinite-dimensional hard core singularity~\cite{Lai1995}.

\paragraph{Random $k$-QSAT---}
We now apply Theorem~\ref{thm:qshearer_qsat} to random $k$-QSAT. 
Random satisfiability has been a workhorse for the study of typical case complexity and heuristic algorithms. 
By tuning the density $\alpha = \frac{M}{N}$ of Erd\H{o}s-R\'enyi-type random interaction graphs, 
 a cornucopia of phases and phase-transitions in the structure of the satisfying space has been discovered.
In the quantum case,  three phases for random $k$-QSAT on qubits and rank-1 projectors are believed to be: 
1) a low density, a product state satisfiable phase (PRODSAT);
2) at intermediate densities for $k$ sufficiently large, a satisfiable phase in which all ground states are entangled (ENTSAT);
and 3) a high density, an unsatisfiable (UNSAT) phase.
The arguably most interesting of these phases, ENTSAT, has been shown to exist using the Quantum Lov\'asz Local Lemma (QLLL)~\cite{AmbainisKS12} for $k \geq 13$.
The application of Theorem 1 to the Erd\H{o}s-R\'enyi (ER) ensemble is not straightforward, 
as the interaction graphs exhibit an unbounded degree distribution, so that a strict application of
Shearer's theorem provides no useful information. 
However, as explained in Appendix~\ref{sec:degree_fluctuations_for_random_qsat}, such local degree fluctuations do not appear to be the source of unsatisfiability in $k$-QSAT for qubits. 

\begin{figure}[htbp] 
\includegraphics[width=276pt]{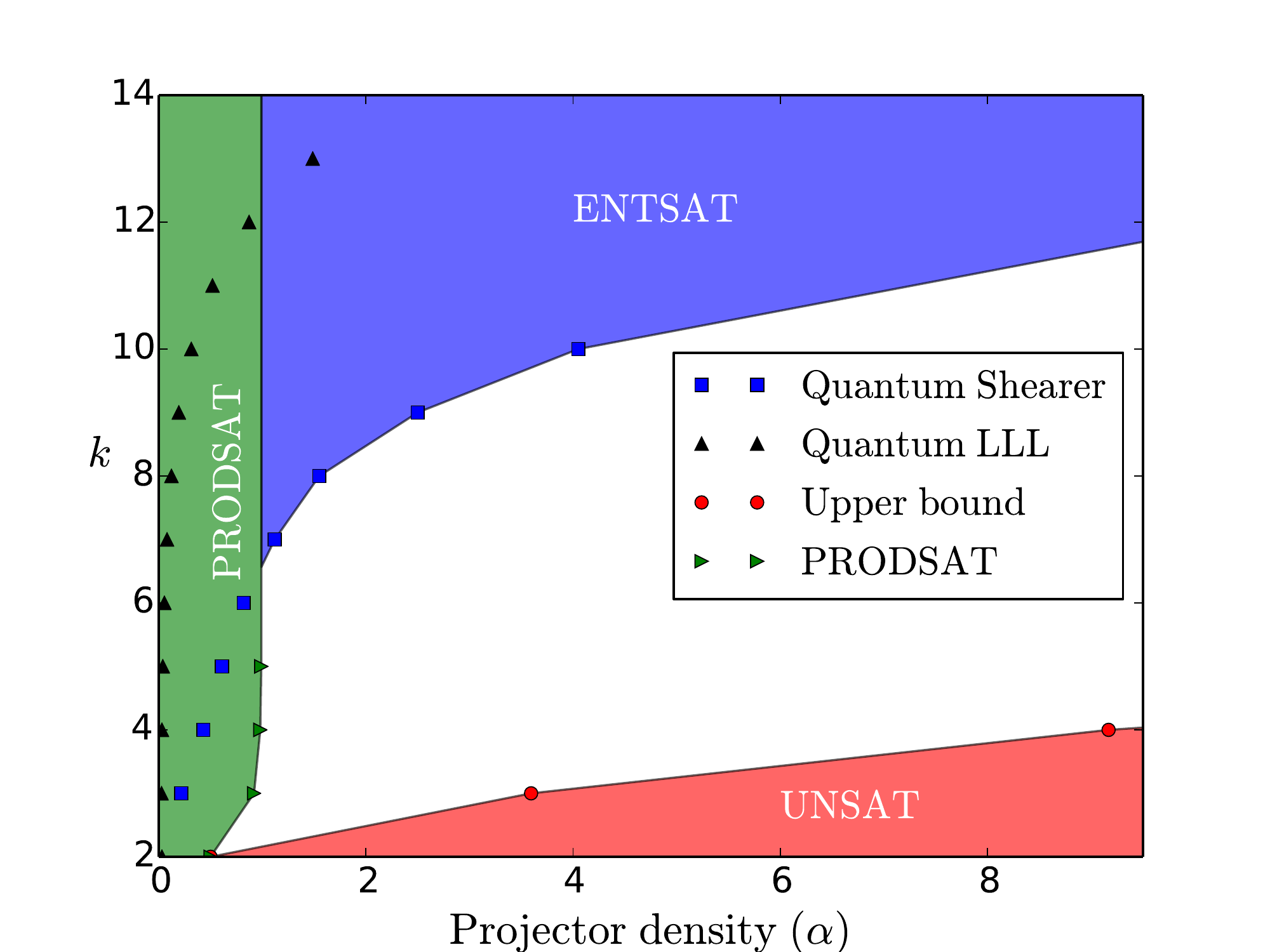}
\protect\caption{\label{fig:randomQSAT_transition}
Our current understanding of the phase diagram
of the random $k$-QSAT. For small $\alpha$, the instances are PRODSAT, with the transition out of PRODSAT approaching $\alpha_P\rightarrow 1$ from below as $k$ grows\cite{PhysRevA.81.062345}. For large $\alpha$, the instances are guaranteed to be UNSAT, with the 
best known upper bound\cite{BravyiMR10} for the satisfiability transition at $\alpha_S^u=0.573\cdot2^{k}$ at $k \geq 4$. In between, there may be an ENTSAT phase if $\alpha_S>\alpha_P$.
The best previous lower bound for the satisfiability transition, $\alpha_S^l$ was given by the Quantum Lovász Local Lemma (QLLL, up triangles)\cite{AmbainisKS12},
and is roughly exponential at large $k$. This work obtains a better lower
bound through the Shearer criterion (squares), lowering the threshold for the existence of an ENTSAT phase from $k=13$ to $k=7$. 
}
\end{figure}

Neglecting this local rare-region effect, we calculate the self-consistent solutions to the disorder-averaged cavity equations using standard methods (population dynamics) \cite{Mezard:2009:IPC:1592967}.
For $\lambda > \lambda_c$, this numerical technique converges to a population of messages which represents the distribution of $q$ and $l$ in the infinite random graph, and from which we can directly estimate the disorder-averaged free energy $f$ and occupation number density $\langle n\rangle = \dfrac{\partial f}{\partial\left(\log\lambda\right)}$. 
The latter is particularly useful, because as $\lambda \to \lambda_c$, we expect $\langle n \rangle$ to exhibit a square root singularity by universality, i.e. $\langle n \rangle \sim (\lambda - \lambda_c)^{1/2}$.
Observing this singularity allows us to estimate $\lambda_c$ accurately and confirms the validity of the numerical approach.
By fitting these singularities, we extract a new and improved (lower) but non-rigorous threshold for the existence of an ENTSAT phase, $k\geq7$, see Fig.~\ref{fig:randomQSAT_transition}. 

\paragraph{Conclusion and open questions---}
By extending the classical Shearer theorem to quantum mechanical systems, we have provided a new statistical mechanical criterion for determining whether a local Hamiltonian is frustration free. 
We have applied this criterion to a large class of regular and random Hamiltonians. 
These instances cover many of the geometries which have been studied in quantum complexity and as parent Hamiltonians for wavefunction-based many-body physics.
In the context of random satisfiability problems, we have provided a set of new lower bounds on the existence of the ENTSAT phase. In particular, these bounds suggest that the ENTSAT phase is eminently more reachable in simulations than previously established.

Theorem 1 depends on the dimension $p$ of the projectors $\Pi_i$ \emph{relative} to the Hilbert space dimension, and not the absolute local dimension $q$ of the qudits. 
At small $q$ (e.g. qubits), there are many Hamiltonians where $R(\ker H)$ is strictly larger than the bound of Theorem 1 (cf. the discussion of random $k$-QSAT). 
Nonetheless, for large $q$ we conjecture that Theorem 1 becomes tight.
This is in sharp contrast to the classical case where the length 4 cycle (periodic chain) already provides a counterexample to the analogous statement \cite{KolipakaS11}. 
However, it is easy to show numerically that this counterexample breaks down for quantum projectors. 

If, indeed, Theorem 1 is tight, there are several striking consequences. The geometrization theorem~\cite{Laumann:2010p7275} states that $\R(\ker H)$ is minimized by almost all choices of the $\Pi_i$'s.
Coupled with tightness, the lattice gas partition function $\ZZ$ then provides a complete characterization of quantum satisfiability for almost all Hamiltonians with large enough qudits.
It also directly lifts the universality of the lattice gas critical exponents to the counting of the ground state entropy of almost all quantum Hamiltonians in the frustration free regime. In this sense, the conjecture amounts to an even larger scope of transferring insights from classical statistical mechanics into the quantum complexity domain. 

While Theorem 1 can guarantee the existence of zero energy states, it does not construct them. 
This is in contrast to classical SAT and commuting QSAT, where efficient constructive classical (under the Lov\'asz and Shearer's assumptions) and quantum algorithms (under the Lov\'asz assumption) are known~\cite{DBLP:journals/jacm/MoserT10,KolipakaS11,schwarz2013information,SattathA15}.

Analogously constructing the wavefunctions corresponding to the solutions of the non-commuting quantum problem would represent a milestone for quantum complexity theory. 

\begin{acknowledgments}
We wish to thank Dorit Aharonov, David Gosset, Antonello Scardicchio, Shivaji Sondhi, Mario Szegedy and Umesh Vazirani for valuable discussions. This work was supported by the ARO Grant W922NF-09-1-0440, NSF Grant CCF-0905626, the DFG grant SFB 1143, ERC Grant 030-8301, and a Sloan Research Fellowship.
\end{acknowledgments}

\bibliographystyle{apsrev4-1} 
\bibliography{ref}

\begin{thebibliography}{70}%
\makeatletter
\providecommand \@ifxundefined [1]{%
 \@ifx{#1\undefined}
}%
\providecommand \@ifnum [1]{%
 \ifnum #1\expandafter \@firstoftwo
 \else \expandafter \@secondoftwo
 \fi
}%
\providecommand \@ifx [1]{%
 \ifx #1\expandafter \@firstoftwo
 \else \expandafter \@secondoftwo
 \fi
}%
\providecommand \natexlab [1]{#1}%
\providecommand \enquote  [1]{``#1''}%
\providecommand \bibnamefont  [1]{#1}%
\providecommand \bibfnamefont [1]{#1}%
\providecommand \citenamefont [1]{#1}%
\providecommand \href@noop [0]{\@secondoftwo}%
\providecommand \href [0]{\begingroup \@sanitize@url \@href}%
\providecommand \@href[1]{\@@startlink{#1}\@@href}%
\providecommand \@@href[1]{\endgroup#1\@@endlink}%
\providecommand \@sanitize@url [0]{\catcode `\\12\catcode `\$12\catcode
  `\&12\catcode `\#12\catcode `\^12\catcode `\_12\catcode `\%12\relax}%
\providecommand \@@startlink[1]{}%
\providecommand \@@endlink[0]{}%
\providecommand \url  [0]{\begingroup\@sanitize@url \@url }%
\providecommand \@url [1]{\endgroup\@href {#1}{\urlprefix }}%
\providecommand \urlprefix  [0]{URL }%
\providecommand \Eprint [0]{\href }%
\providecommand \doibase [0]{http://dx.doi.org/}%
\providecommand \selectlanguage [0]{\@gobble}%
\providecommand \bibinfo  [0]{\@secondoftwo}%
\providecommand \bibfield  [0]{\@secondoftwo}%
\providecommand \translation [1]{[#1]}%
\providecommand \BibitemOpen [0]{}%
\providecommand \bibitemStop [0]{}%
\providecommand \bibitemNoStop [0]{.\EOS\space}%
\providecommand \EOS [0]{\spacefactor3000\relax}%
\providecommand \BibitemShut  [1]{\csname bibitem#1\endcsname}%
\let\auto@bib@innerbib\@empty
\bibitem [{\citenamefont {Kitaev}(2003)}]{Kitaev2003}%
  \BibitemOpen
  \bibfield  {author} {\bibinfo {author} {\bibfnamefont {A.}~\bibnamefont
  {Kitaev}},\ }\href {\doibase http://dx.doi.org/10.1016/S0003-4916(02)00018-0}
  {\bibfield  {journal} {\bibinfo  {journal} {Annals of Physics}\ }\textbf
  {\bibinfo {volume} {303}},\ \bibinfo {pages} {2 } (\bibinfo {year}
  {2003})}\BibitemShut {NoStop}%
\bibitem [{\citenamefont {Gottesman}(2010)}]{Gottesman2010}%
  \BibitemOpen
  \bibfield  {author} {\bibinfo {author} {\bibfnamefont {D.}~\bibnamefont
  {Gottesman}},\ }in\ \href {\doibase 10.1090/psapm/068/2762145} {\emph
  {\bibinfo {booktitle} {Quantum information science and its contributions to
  mathematics}}},\ \bibinfo {series} {Proc. Sympos. Appl. Math.}, Vol.~\bibinfo
  {volume} {68}\ (\bibinfo  {publisher} {Amer. Math. Soc., Providence, RI},\
  \bibinfo {year} {2010})\ pp.\ \bibinfo {pages} {13--58}\BibitemShut {NoStop}%
\bibitem [{\citenamefont {Levin}\ and\ \citenamefont {Wen}(2005)}]{LevinWen}%
  \BibitemOpen
  \bibfield  {author} {\bibinfo {author} {\bibfnamefont {M.~A.}\ \bibnamefont
  {Levin}}\ and\ \bibinfo {author} {\bibfnamefont {X.-G.}\ \bibnamefont
  {Wen}},\ }\href {\doibase 10.1103/PhysRevB.71.045110} {\bibfield  {journal}
  {\bibinfo  {journal} {Phys. Rev. B}\ }\textbf {\bibinfo {volume} {71}},\
  \bibinfo {pages} {045110} (\bibinfo {year} {2005})}\BibitemShut {NoStop}%
\bibitem [{\citenamefont {Affleck}\ \emph {et~al.}(1987)\citenamefont
  {Affleck}, \citenamefont {Kennedy}, \citenamefont {Lieb},\ and\ \citenamefont
  {Tasaki}}]{AKLT1987}%
  \BibitemOpen
  \bibfield  {author} {\bibinfo {author} {\bibfnamefont {I.}~\bibnamefont
  {Affleck}}, \bibinfo {author} {\bibfnamefont {T.}~\bibnamefont {Kennedy}},
  \bibinfo {author} {\bibfnamefont {E.~H.}\ \bibnamefont {Lieb}}, \ and\
  \bibinfo {author} {\bibfnamefont {H.}~\bibnamefont {Tasaki}},\ }\href
  {\doibase 10.1103/PhysRevLett.59.799} {\bibfield  {journal} {\bibinfo
  {journal} {Phys. Rev. Lett.}\ }\textbf {\bibinfo {volume} {59}},\ \bibinfo
  {pages} {799} (\bibinfo {year} {1987})}\BibitemShut {NoStop}%
\bibitem [{\citenamefont {Rokhsar}\ and\ \citenamefont
  {Kivelson}(1988)}]{RKOriginal}%
  \BibitemOpen
  \bibfield  {author} {\bibinfo {author} {\bibfnamefont {D.~S.}\ \bibnamefont
  {Rokhsar}}\ and\ \bibinfo {author} {\bibfnamefont {S.~A.}\ \bibnamefont
  {Kivelson}},\ }\href {\doibase 10.1103/PhysRevLett.61.2376} {\bibfield
  {journal} {\bibinfo  {journal} {Phys. Rev. Lett.}\ }\textbf {\bibinfo
  {volume} {61}},\ \bibinfo {pages} {2376} (\bibinfo {year}
  {1988})}\BibitemShut {NoStop}%
\bibitem [{\citenamefont {Castelnovo}\ \emph {et~al.}(2005)\citenamefont
  {Castelnovo}, \citenamefont {Chamon}, \citenamefont {Mudry},\ and\
  \citenamefont {Pujol}}]{CastelnovoRK}%
  \BibitemOpen
  \bibfield  {author} {\bibinfo {author} {\bibfnamefont {C.}~\bibnamefont
  {Castelnovo}}, \bibinfo {author} {\bibfnamefont {C.}~\bibnamefont {Chamon}},
  \bibinfo {author} {\bibfnamefont {C.}~\bibnamefont {Mudry}}, \ and\ \bibinfo
  {author} {\bibfnamefont {P.}~\bibnamefont {Pujol}},\ }\href {\doibase
  http://dx.doi.org/10.1016/j.aop.2005.01.006} {\bibfield  {journal} {\bibinfo
  {journal} {Annals of Physics}\ }\textbf {\bibinfo {volume} {318}},\ \bibinfo
  {pages} {316 } (\bibinfo {year} {2005})}\BibitemShut {NoStop}%
\bibitem [{\citenamefont {Perez-Garcia}\ \emph {et~al.}(2008)\citenamefont
  {Perez-Garcia}, \citenamefont {Verstraete}, \citenamefont {Wolf},\ and\
  \citenamefont {Cirac}}]{perez2008peps}%
  \BibitemOpen
  \bibfield  {author} {\bibinfo {author} {\bibfnamefont {D.}~\bibnamefont
  {Perez-Garcia}}, \bibinfo {author} {\bibfnamefont {F.}~\bibnamefont
  {Verstraete}}, \bibinfo {author} {\bibfnamefont {M.~M.}\ \bibnamefont
  {Wolf}}, \ and\ \bibinfo {author} {\bibfnamefont {J.~I.}\ \bibnamefont
  {Cirac}},\ }\href@noop {} {\bibfield  {journal} {\bibinfo  {journal} {Quantum
  Information \& Computation}\ }\textbf {\bibinfo {volume} {8}},\ \bibinfo
  {pages} {650} (\bibinfo {year} {2008})}\BibitemShut {NoStop}%
\bibitem [{\citenamefont {Bravyi}(2011)}]{bravyi2011efficient}%
  \BibitemOpen
  \bibfield  {author} {\bibinfo {author} {\bibfnamefont {S.}~\bibnamefont
  {Bravyi}},\ }in\ \href
  {http://www.ams.org/books/conm/536/10552/conm536-10552.pdf} {\emph {\bibinfo
  {booktitle} {Contemporary Mathematics}}},\ Vol.\ \bibinfo {volume} {536},\
  \bibinfo {editor} {edited by\ \bibinfo {editor} {\bibfnamefont
  {K.}~\bibnamefont {Mahdavi}}, \bibinfo {editor} {\bibfnamefont
  {D.}~\bibnamefont {Koslover}}, \ and\ \bibinfo {editor} {\bibfnamefont
  {L.~L.}\ \bibnamefont {Brown}}}\ (\bibinfo  {publisher} {American
  Mathematical Society},\ \bibinfo {year} {2011})\BibitemShut {NoStop}%
\bibitem [{\citenamefont {Gosset}\ and\ \citenamefont
  {Nagaj}(2013)}]{GossetN13}%
  \BibitemOpen
  \bibfield  {author} {\bibinfo {author} {\bibfnamefont {D.}~\bibnamefont
  {Gosset}}\ and\ \bibinfo {author} {\bibfnamefont {D.}~\bibnamefont {Nagaj}},\
  }in\  \cite{DBLP:conf/focs/2013},\ pp.\ \bibinfo {pages}
  {756--765}\BibitemShut {NoStop}%
\bibitem [{\citenamefont {Aharonov}\ \emph {et~al.}(2007)\citenamefont
  {Aharonov}, \citenamefont {Gottesman}, \citenamefont {Irani},\ and\
  \citenamefont {Kempe}}]{aharonov_power_2007}%
  \BibitemOpen
  \bibfield  {author} {\bibinfo {author} {\bibfnamefont {D.}~\bibnamefont
  {Aharonov}}, \bibinfo {author} {\bibfnamefont {D.}~\bibnamefont {Gottesman}},
  \bibinfo {author} {\bibfnamefont {S.}~\bibnamefont {Irani}}, \ and\ \bibinfo
  {author} {\bibfnamefont {J.}~\bibnamefont {Kempe}},\ }in\ \href
  {http://portal.acm.org/citation.cfm?id=1333875.1334210} {\emph {\bibinfo
  {booktitle} {Proceedings of the 48th Annual {IEEE} Symposium on Foundations
  of Computer Science}}}\ (\bibinfo  {publisher} {{IEEE} Computer Society},\
  \bibinfo {year} {2007})\ pp.\ \bibinfo {pages} {373--383}\BibitemShut
  {NoStop}%
\bibitem [{\citenamefont {Shearer}(1985)}]{Shearer1985}%
  \BibitemOpen
  \bibfield  {author} {\bibinfo {author} {\bibfnamefont {J.}~\bibnamefont
  {Shearer}},\ }\href {\doibase 10.1007/BF02579368} {\bibfield  {journal}
  {\bibinfo  {journal} {Combinatorica}\ }\textbf {\bibinfo {volume} {5}},\
  \bibinfo {pages} {241} (\bibinfo {year} {1985})}\BibitemShut {NoStop}%
\bibitem [{\citenamefont {Achlioptas}\ and\ \citenamefont
  {Peres}(2004)}]{achlioptas03threshold}%
  \BibitemOpen
  \bibfield  {author} {\bibinfo {author} {\bibfnamefont {D.}~\bibnamefont
  {Achlioptas}}\ and\ \bibinfo {author} {\bibfnamefont {Y.}~\bibnamefont
  {Peres}},\ }\href {\doibase 10.1090/S0894-0347-04-00464-3} {\bibfield
  {journal} {\bibinfo  {journal} {J. Amer. Math. Soc.}\ }\textbf {\bibinfo
  {volume} {17}},\ \bibinfo {pages} {947} (\bibinfo {year} {2004})}\BibitemShut
  {NoStop}%
\bibitem [{\citenamefont {Rathi}\ \emph {et~al.}(2010)\citenamefont {Rathi},
  \citenamefont {Aurell}, \citenamefont {Rasmussen},\ and\ \citenamefont
  {Skoglund}}]{rathi10bounds}%
  \BibitemOpen
  \bibfield  {author} {\bibinfo {author} {\bibfnamefont {V.}~\bibnamefont
  {Rathi}}, \bibinfo {author} {\bibfnamefont {E.}~\bibnamefont {Aurell}},
  \bibinfo {author} {\bibfnamefont {L.~K.}\ \bibnamefont {Rasmussen}}, \ and\
  \bibinfo {author} {\bibfnamefont {M.}~\bibnamefont {Skoglund}},\ }in\
  \cite{DBLP:conf/sat/2010},\ pp.\ \bibinfo {pages} {264--277}\BibitemShut
  {NoStop}%
\bibitem [{Note1()}]{Note1}%
  \BibitemOpen
  \bibinfo {note} {This may always be done by shifting and deforming the
  eigenvalues of the local Hamiltonian terms without influencing the
  frustration free ground state space.}\BibitemShut {Stop}%
\bibitem [{\citenamefont {Scott}\ and\ \citenamefont
  {Sokal}(2005)}]{Scott2005}%
  \BibitemOpen
  \bibfield  {author} {\bibinfo {author} {\bibfnamefont {A.}~\bibnamefont
  {Scott}}\ and\ \bibinfo {author} {\bibfnamefont {A.}~\bibnamefont {Sokal}},\
  }\href {\doibase 10.1007/s10955-004-2055-4} {\bibfield  {journal} {\bibinfo
  {journal} {Journal of Statistical Physics}\ }\textbf {\bibinfo {volume}
  {118}},\ \bibinfo {pages} {1151} (\bibinfo {year} {2005})}\BibitemShut
  {NoStop}%
\bibitem [{\citenamefont {Hoffmann}(2010{\natexlab{a}})}]{Hoffmann2010}%
  \BibitemOpen
  \bibfield  {author} {\bibinfo {author} {\bibfnamefont {C.}~\bibnamefont
  {Hoffmann}},\ }\href@noop {} {\bibfield  {journal} {\bibinfo  {journal}
  {Ph.D. Thesis Universit\"{a}t des Saarlandes}\ } (\bibinfo {year}
  {2010}{\natexlab{a}})}\BibitemShut {NoStop}%
\bibitem [{\citenamefont {Groeneveld}(1962)}]{Groeneveld}%
  \BibitemOpen
  \bibfield  {author} {\bibinfo {author} {\bibfnamefont {J.}~\bibnamefont
  {Groeneveld}},\ }\href
  {http://www.sciencedirect.com/science/article/pii/0031916362901981}
  {\bibfield  {journal} {\bibinfo  {journal} {Phys. Lett.}\ }\textbf {\bibinfo
  {volume} {3}},\ \bibinfo {pages} {50} (\bibinfo {year} {1962})}\BibitemShut
  {NoStop}%
\bibitem [{\citenamefont {Gaunt}\ and\ \citenamefont
  {Fisher}(1965)}]{Gaunt1965}%
  \BibitemOpen
  \bibfield  {author} {\bibinfo {author} {\bibfnamefont {D.~S.}\ \bibnamefont
  {Gaunt}}\ and\ \bibinfo {author} {\bibfnamefont {M.~E.}\ \bibnamefont
  {Fisher}},\ }\href {\doibase http://dx.doi.org/10.1063/1.1697217} {\bibfield
  {journal} {\bibinfo  {journal} {The Journal of Chemical Physics}\ }\textbf
  {\bibinfo {volume} {43}},\ \bibinfo {pages} {2840} (\bibinfo {year}
  {1965})}\BibitemShut {NoStop}%
\bibitem [{\citenamefont {Gaunt}(1967)}]{Gaunt1967}%
  \BibitemOpen
  \bibfield  {author} {\bibinfo {author} {\bibfnamefont {D.~S.}\ \bibnamefont
  {Gaunt}},\ }\href {\doibase http://dx.doi.org/10.1063/1.1841195} {\bibfield
  {journal} {\bibinfo  {journal} {The Journal of Chemical Physics}\ }\textbf
  {\bibinfo {volume} {46}},\ \bibinfo {pages} {3237} (\bibinfo {year}
  {1967})}\BibitemShut {NoStop}%
\bibitem [{\citenamefont {Heilmann}\ and\ \citenamefont
  {Lieb}(1972)}]{heilmann1972}%
  \BibitemOpen
  \bibfield  {author} {\bibinfo {author} {\bibfnamefont {O.~J.}\ \bibnamefont
  {Heilmann}}\ and\ \bibinfo {author} {\bibfnamefont {E.~H.}\ \bibnamefont
  {Lieb}},\ }\href {http://projecteuclid.org/euclid.cmp/1103857921} {\bibfield
  {journal} {\bibinfo  {journal} {Comm. Math. Phys.}\ }\textbf {\bibinfo
  {volume} {25}},\ \bibinfo {pages} {190} (\bibinfo {year} {1972})}\BibitemShut
  {NoStop}%
\bibitem [{\citenamefont {Baxter}(1980)}]{Baxter1980}%
  \BibitemOpen
  \bibfield  {author} {\bibinfo {author} {\bibfnamefont {R.~J.}\ \bibnamefont
  {Baxter}},\ }\href {http://stacks.iop.org/0305-4470/13/i=3/a=007} {\bibfield
  {journal} {\bibinfo  {journal} {Journal of Physics A: Mathematical and
  General}\ }\textbf {\bibinfo {volume} {13}},\ \bibinfo {pages} {L61}
  (\bibinfo {year} {1980})}\BibitemShut {NoStop}%
\bibitem [{\citenamefont {Poland}(1984)}]{Poland1984}%
  \BibitemOpen
  \bibfield  {author} {\bibinfo {author} {\bibfnamefont {D.}~\bibnamefont
  {Poland}},\ }\href {\doibase 10.1007/BF01014388} {\bibfield  {journal}
  {\bibinfo  {journal} {Journal of Statistical Physics}\ }\textbf {\bibinfo
  {volume} {35}},\ \bibinfo {pages} {341} (\bibinfo {year} {1984})}\BibitemShut
  {NoStop}%
\bibitem [{\citenamefont {Baram}\ and\ \citenamefont
  {Luban}(1987)}]{PhysRevA.36.760}%
  \BibitemOpen
  \bibfield  {author} {\bibinfo {author} {\bibfnamefont {A.}~\bibnamefont
  {Baram}}\ and\ \bibinfo {author} {\bibfnamefont {M.}~\bibnamefont {Luban}},\
  }\href {\doibase 10.1103/PhysRevA.36.760} {\bibfield  {journal} {\bibinfo
  {journal} {Phys. Rev. A}\ }\textbf {\bibinfo {volume} {36}},\ \bibinfo
  {pages} {760} (\bibinfo {year} {1987})}\BibitemShut {NoStop}%
\bibitem [{\citenamefont {Todo}(1999)}]{todo1999transfer}%
  \BibitemOpen
  \bibfield  {author} {\bibinfo {author} {\bibfnamefont {S.}~\bibnamefont
  {Todo}},\ }\href {\doibase 10.1142/S0129183199000401} {\bibfield  {journal}
  {\bibinfo  {journal} {Internat. J. Modern Phys. C}\ }\textbf {\bibinfo
  {volume} {10}},\ \bibinfo {pages} {517} (\bibinfo {year} {1999})}\BibitemShut
  {NoStop}%
\bibitem [{\citenamefont {Lai}\ and\ \citenamefont {Fisher}(1995)}]{Lai1995}%
  \BibitemOpen
  \bibfield  {author} {\bibinfo {author} {\bibfnamefont {S.-N.}\ \bibnamefont
  {Lai}}\ and\ \bibinfo {author} {\bibfnamefont {M.~E.}\ \bibnamefont
  {Fisher}},\ }\href {\doibase http://dx.doi.org/10.1063/1.470178} {\bibfield
  {journal} {\bibinfo  {journal} {The Journal of Chemical Physics}\ }\textbf
  {\bibinfo {volume} {103}},\ \bibinfo {pages} {8144} (\bibinfo {year}
  {1995})}\BibitemShut {NoStop}%
\bibitem [{\citenamefont {Park}\ and\ \citenamefont
  {Fisher}(1999)}]{PhysRevE.60.6323}%
  \BibitemOpen
  \bibfield  {author} {\bibinfo {author} {\bibfnamefont {Y.}~\bibnamefont
  {Park}}\ and\ \bibinfo {author} {\bibfnamefont {M.~E.}\ \bibnamefont
  {Fisher}},\ }\href {\doibase 10.1103/PhysRevE.60.6323} {\bibfield  {journal}
  {\bibinfo  {journal} {Phys. Rev. E}\ }\textbf {\bibinfo {volume} {60}},\
  \bibinfo {pages} {6323} (\bibinfo {year} {1999})}\BibitemShut {NoStop}%
\bibitem [{\citenamefont {Movassagh}\ \emph {et~al.}(2010)\citenamefont
  {Movassagh}, \citenamefont {Farhi}, \citenamefont {Goldstone}, \citenamefont
  {Nagaj}, \citenamefont {Osborne},\ and\ \citenamefont
  {Shor}}]{movassagh2010unfrustrated}%
  \BibitemOpen
  \bibfield  {author} {\bibinfo {author} {\bibfnamefont {R.}~\bibnamefont
  {Movassagh}}, \bibinfo {author} {\bibfnamefont {E.}~\bibnamefont {Farhi}},
  \bibinfo {author} {\bibfnamefont {J.}~\bibnamefont {Goldstone}}, \bibinfo
  {author} {\bibfnamefont {D.}~\bibnamefont {Nagaj}}, \bibinfo {author}
  {\bibfnamefont {T.~J.}\ \bibnamefont {Osborne}}, \ and\ \bibinfo {author}
  {\bibfnamefont {P.~W.}\ \bibnamefont {Shor}},\ }\href {\doibase
  10.1103/PhysRevA.82.012318} {\bibfield  {journal} {\bibinfo  {journal} {Phys.
  Rev. A}\ }\textbf {\bibinfo {volume} {82}},\ \bibinfo {pages} {012318}
  (\bibinfo {year} {2010})}\BibitemShut {NoStop}%
\bibitem [{\citenamefont {Coudron}\ and\ \citenamefont
  {Movassagh}(2012)}]{coudron2012unfrustration}%
  \BibitemOpen
  \bibfield  {author} {\bibinfo {author} {\bibfnamefont {M.}~\bibnamefont
  {Coudron}}\ and\ \bibinfo {author} {\bibfnamefont {R.}~\bibnamefont
  {Movassagh}},\ }\href {http://arxiv.org/pdf/1209.4395v2.pdf} {\bibfield
  {journal} {\bibinfo  {journal} {arXiv preprint arXiv:1209.4395}\ } (\bibinfo
  {year} {2012})}\BibitemShut {NoStop}%
\bibitem [{\citenamefont {Yang}\ and\ \citenamefont
  {Lee}(1952)}]{PhysRev.87.404}%
  \BibitemOpen
  \bibfield  {author} {\bibinfo {author} {\bibfnamefont {C.~N.}\ \bibnamefont
  {Yang}}\ and\ \bibinfo {author} {\bibfnamefont {T.~D.}\ \bibnamefont {Lee}},\
  }\href {\doibase 10.1103/PhysRev.87.404} {\bibfield  {journal} {\bibinfo
  {journal} {Phys. Rev.}\ }\textbf {\bibinfo {volume} {87}},\ \bibinfo {pages}
  {404} (\bibinfo {year} {1952})}\BibitemShut {NoStop}%
\bibitem [{\citenamefont {Lee}\ and\ \citenamefont
  {Yang}(1952)}]{PhysRev.87.410}%
  \BibitemOpen
  \bibfield  {author} {\bibinfo {author} {\bibfnamefont {T.~D.}\ \bibnamefont
  {Lee}}\ and\ \bibinfo {author} {\bibfnamefont {C.~N.}\ \bibnamefont {Yang}},\
  }\href {\doibase 10.1103/PhysRev.87.410} {\bibfield  {journal} {\bibinfo
  {journal} {Phys. Rev.}\ }\textbf {\bibinfo {volume} {87}},\ \bibinfo {pages}
  {410} (\bibinfo {year} {1952})}\BibitemShut {NoStop}%
\bibitem [{\citenamefont {Shapir}(1982)}]{Shapir1982}%
  \BibitemOpen
  \bibfield  {author} {\bibinfo {author} {\bibfnamefont {Y.}~\bibnamefont
  {Shapir}},\ }\href {http://stacks.iop.org/0305-4470/15/i=8/a=010} {\bibfield
  {journal} {\bibinfo  {journal} {Journal of Physics A: Mathematical and
  General}\ }\textbf {\bibinfo {volume} {15}},\ \bibinfo {pages} {L433}
  (\bibinfo {year} {1982})}\BibitemShut {NoStop}%
\bibitem [{\citenamefont {Fisher}(1978)}]{PhysRevLett.40.1610}%
  \BibitemOpen
  \bibfield  {author} {\bibinfo {author} {\bibfnamefont {M.~E.}\ \bibnamefont
  {Fisher}},\ }\href {\doibase 10.1103/PhysRevLett.40.1610} {\bibfield
  {journal} {\bibinfo  {journal} {Phys. Rev. Lett.}\ }\textbf {\bibinfo
  {volume} {40}},\ \bibinfo {pages} {1610} (\bibinfo {year}
  {1978})}\BibitemShut {NoStop}%
\bibitem [{\citenamefont {Cardy}(1985)}]{PhysRevLett.54.1354}%
  \BibitemOpen
  \bibfield  {author} {\bibinfo {author} {\bibfnamefont {J.~L.}\ \bibnamefont
  {Cardy}},\ }\href {\doibase 10.1103/PhysRevLett.54.1354} {\bibfield
  {journal} {\bibinfo  {journal} {Phys. Rev. Lett.}\ }\textbf {\bibinfo
  {volume} {54}},\ \bibinfo {pages} {1354} (\bibinfo {year}
  {1985})}\BibitemShut {NoStop}%
\bibitem [{\citenamefont {Mezard}\ and\ \citenamefont
  {Parisi}(2001)}]{Mezard2001}%
  \BibitemOpen
  \bibfield  {author} {\bibinfo {author} {\bibfnamefont {M.}~\bibnamefont
  {Mezard}}\ and\ \bibinfo {author} {\bibfnamefont {G.}~\bibnamefont
  {Parisi}},\ }\href {\doibase 10.1007/PL00011099} {\bibfield  {journal}
  {\bibinfo  {journal} {The European Physical Journal B - Condensed Matter and
  Complex Systems}\ }\textbf {\bibinfo {volume} {20}},\ \bibinfo {pages} {217}
  (\bibinfo {year} {2001})}\BibitemShut {NoStop}%
\bibitem [{\citenamefont {Laumann}\ \emph {et~al.}(2008)\citenamefont
  {Laumann}, \citenamefont {Scardicchio},\ and\ \citenamefont
  {Sondhi}}]{laumannqcav}%
  \BibitemOpen
  \bibfield  {author} {\bibinfo {author} {\bibfnamefont {C.}~\bibnamefont
  {Laumann}}, \bibinfo {author} {\bibfnamefont {A.}~\bibnamefont
  {Scardicchio}}, \ and\ \bibinfo {author} {\bibfnamefont {S.~L.}\ \bibnamefont
  {Sondhi}},\ }\href {\doibase 10.1103/PhysRevB.78.134424} {\bibfield
  {journal} {\bibinfo  {journal} {Phys. Rev. B}\ }\textbf {\bibinfo {volume}
  {78}},\ \bibinfo {pages} {134424} (\bibinfo {year} {2008})}\BibitemShut
  {NoStop}%
\bibitem [{\citenamefont {Leifer}\ and\ \citenamefont
  {Poulin}(2008)}]{poulinqcav}%
  \BibitemOpen
  \bibfield  {author} {\bibinfo {author} {\bibfnamefont {M.}~\bibnamefont
  {Leifer}}\ and\ \bibinfo {author} {\bibfnamefont {D.}~\bibnamefont
  {Poulin}},\ }\href {\doibase http://dx.doi.org/10.1016/j.aop.2007.10.001}
  {\bibfield  {journal} {\bibinfo  {journal} {Annals of Physics}\ }\textbf
  {\bibinfo {volume} {323}},\ \bibinfo {pages} {1899 } (\bibinfo {year}
  {2008})}\BibitemShut {NoStop}%
\bibitem [{\citenamefont {Hastings}(2007)}]{hastingsqcav}%
  \BibitemOpen
  \bibfield  {author} {\bibinfo {author} {\bibfnamefont {M.~B.}\ \bibnamefont
  {Hastings}},\ }\href {\doibase 10.1103/PhysRevB.76.201102} {\bibfield
  {journal} {\bibinfo  {journal} {Phys. Rev. B}\ }\textbf {\bibinfo {volume}
  {76}},\ \bibinfo {pages} {201102} (\bibinfo {year} {2007})}\BibitemShut
  {NoStop}%
\bibitem [{\citenamefont {Mezard}\ and\ \citenamefont
  {Montanari}(2009)}]{Mezard:2009:IPC:1592967}%
  \BibitemOpen
  \bibfield  {author} {\bibinfo {author} {\bibfnamefont {M.}~\bibnamefont
  {Mezard}}\ and\ \bibinfo {author} {\bibfnamefont {A.}~\bibnamefont
  {Montanari}},\ }\href@noop {} {\emph {\bibinfo {title} {Information, Physics,
  and Computation}}}\ (\bibinfo  {publisher} {Oxford University Press, Inc.},\
  \bibinfo {address} {New York, NY, USA},\ \bibinfo {year} {2009})\BibitemShut
  {NoStop}%
\bibitem [{\citenamefont {Zdeborov{\'a}}\ and\ \citenamefont
  {M{\'e}zard}(2006)}]{lenkamatchings}%
  \BibitemOpen
  \bibfield  {author} {\bibinfo {author} {\bibfnamefont {L.}~\bibnamefont
  {Zdeborov{\'a}}}\ and\ \bibinfo {author} {\bibfnamefont {M.}~\bibnamefont
  {M{\'e}zard}},\ }\href {http://stacks.iop.org/1742-5468/2006/i=05/a=P05003}
  {\bibfield  {journal} {\bibinfo  {journal} {Journal of Statistical Mechanics:
  Theory and Experiment}\ }\textbf {\bibinfo {volume} {2006}},\ \bibinfo
  {pages} {P05003} (\bibinfo {year} {2006})}\BibitemShut {NoStop}%
\bibitem [{\citenamefont {Laumann}\ \emph
  {et~al.}(2010{\natexlab{a}})\citenamefont {Laumann}, \citenamefont
  {Moessner}, \citenamefont {Scarddichio},\ and\ \citenamefont
  {Sondhi}}]{Laumann:2010p7275}%
  \BibitemOpen
  \bibfield  {author} {\bibinfo {author} {\bibfnamefont {C.~R.}\ \bibnamefont
  {Laumann}}, \bibinfo {author} {\bibfnamefont {R.}~\bibnamefont {Moessner}},
  \bibinfo {author} {\bibfnamefont {A.}~\bibnamefont {Scarddichio}}, \ and\
  \bibinfo {author} {\bibfnamefont {S.~L.}\ \bibnamefont {Sondhi}},\ }\href
  {http://www.rintonpress.com/xxqic10/qic-10-12/0001-0015.pdf} {\bibfield
  {journal} {\bibinfo  {journal} {Quantum Information {\&} Computation}\
  }\textbf {\bibinfo {volume} {10}},\ \bibinfo {pages} {1} (\bibinfo {year}
  {2010}{\natexlab{a}})}\BibitemShut {NoStop}%
\bibitem [{\citenamefont {Ambainis}\ \emph {et~al.}(2012)\citenamefont
  {Ambainis}, \citenamefont {Kempe},\ and\ \citenamefont
  {Sattath}}]{AmbainisKS12}%
  \BibitemOpen
  \bibfield  {author} {\bibinfo {author} {\bibfnamefont {A.}~\bibnamefont
  {Ambainis}}, \bibinfo {author} {\bibfnamefont {J.}~\bibnamefont {Kempe}}, \
  and\ \bibinfo {author} {\bibfnamefont {O.}~\bibnamefont {Sattath}},\ }\href
  {\doibase 10.1145/2371656.2371659} {\bibfield  {journal} {\bibinfo  {journal}
  {J. {ACM}}\ }\textbf {\bibinfo {volume} {59}},\ \bibinfo {pages} {24}
  (\bibinfo {year} {2012})}\BibitemShut {NoStop}%
\bibitem [{\citenamefont {Laumann}\ \emph
  {et~al.}(2010{\natexlab{b}})\citenamefont {Laumann}, \citenamefont
  {L\"auchli}, \citenamefont {Moessner}, \citenamefont {Scardicchio},\ and\
  \citenamefont {Sondhi}}]{PhysRevA.81.062345}%
  \BibitemOpen
  \bibfield  {author} {\bibinfo {author} {\bibfnamefont {C.~R.}\ \bibnamefont
  {Laumann}}, \bibinfo {author} {\bibfnamefont {A.~M.}\ \bibnamefont
  {L\"auchli}}, \bibinfo {author} {\bibfnamefont {R.}~\bibnamefont {Moessner}},
  \bibinfo {author} {\bibfnamefont {A.}~\bibnamefont {Scardicchio}}, \ and\
  \bibinfo {author} {\bibfnamefont {S.~L.}\ \bibnamefont {Sondhi}},\ }\href
  {\doibase 10.1103/PhysRevA.81.062345} {\bibfield  {journal} {\bibinfo
  {journal} {Phys. Rev. A}\ }\textbf {\bibinfo {volume} {81}},\ \bibinfo
  {pages} {062345} (\bibinfo {year} {2010}{\natexlab{b}})}\BibitemShut
  {NoStop}%
\bibitem [{\citenamefont {Bravyi}\ \emph {et~al.}(2010)\citenamefont {Bravyi},
  \citenamefont {Moore},\ and\ \citenamefont {Russell}}]{BravyiMR10}%
  \BibitemOpen
  \bibfield  {author} {\bibinfo {author} {\bibfnamefont {S.}~\bibnamefont
  {Bravyi}}, \bibinfo {author} {\bibfnamefont {C.}~\bibnamefont {Moore}}, \
  and\ \bibinfo {author} {\bibfnamefont {A.}~\bibnamefont {Russell}},\ }in\
  \href {http://conference.itcs.tsinghua.edu.cn/ICS2010/content/papers/37.html}
  {\emph {\bibinfo {booktitle} {Innovations in Computer Science}}}\ (\bibinfo
  {year} {2010})\ pp.\ \bibinfo {pages} {482--489}\BibitemShut {NoStop}%
\bibitem [{\citenamefont {Kolipaka}\ and\ \citenamefont
  {Szegedy}(2011)}]{KolipakaS11}%
  \BibitemOpen
  \bibfield  {author} {\bibinfo {author} {\bibfnamefont {K.~B.~R.}\
  \bibnamefont {Kolipaka}}\ and\ \bibinfo {author} {\bibfnamefont
  {M.}~\bibnamefont {Szegedy}},\ }in\ \href {\doibase 10.1145/1993636.1993669}
  {\emph {\bibinfo {booktitle} {Proceedings of the 43rd {ACM} Symposium on
  Theory of Computing, {STOC} 2011}}}\ (\bibinfo {year} {2011})\ pp.\ \bibinfo
  {pages} {235--244}\BibitemShut {NoStop}%
\bibitem [{\citenamefont {Moser}\ and\ \citenamefont
  {Tardos}(2010)}]{DBLP:journals/jacm/MoserT10}%
  \BibitemOpen
  \bibfield  {author} {\bibinfo {author} {\bibfnamefont {R.~A.}\ \bibnamefont
  {Moser}}\ and\ \bibinfo {author} {\bibfnamefont {G.}~\bibnamefont {Tardos}},\
  }\href {\doibase 10.1145/1667053.1667060} {\bibfield  {journal} {\bibinfo
  {journal} {J. {ACM}}\ }\textbf {\bibinfo {volume} {57}} (\bibinfo {year}
  {2010}),\ 10.1145/1667053.1667060}\BibitemShut {NoStop}%
\bibitem [{\citenamefont {Schwarz}\ \emph {et~al.}(2013)\citenamefont
  {Schwarz}, \citenamefont {Cubitt},\ and\ \citenamefont
  {Verstraete}}]{schwarz2013information}%
  \BibitemOpen
  \bibfield  {author} {\bibinfo {author} {\bibfnamefont {M.}~\bibnamefont
  {Schwarz}}, \bibinfo {author} {\bibfnamefont {T.~S.}\ \bibnamefont {Cubitt}},
  \ and\ \bibinfo {author} {\bibfnamefont {F.}~\bibnamefont {Verstraete}},\
  }\href {http://arxiv.org/pdf/1311.6474v1.pdf} {\bibfield  {journal} {\bibinfo
   {journal} {arXiv:1311.6474}\ } (\bibinfo {year} {2013})}\BibitemShut
  {NoStop}%
\bibitem [{\citenamefont {Sattath}\ and\ \citenamefont
  {Arad}(2015)}]{SattathA15}%
  \BibitemOpen
  \bibfield  {author} {\bibinfo {author} {\bibfnamefont {O.}~\bibnamefont
  {Sattath}}\ and\ \bibinfo {author} {\bibfnamefont {I.}~\bibnamefont {Arad}},\
  }\href {http://www.rintonpress.com/xxqic15/qic-15-1112/0987-0996.pdf}
  {\bibfield  {journal} {\bibinfo  {journal} {Quantum Information {\&}
  Computation}\ }\textbf {\bibinfo {volume} {15}},\ \bibinfo {pages} {987}
  (\bibinfo {year} {2015})}\BibitemShut {NoStop}%
\bibitem [{DBL(2013)}]{DBLP:conf/focs/2013}%
  \BibitemOpen
  \href {http://ieeexplore.ieee.org/xpl/mostRecentIssue.jsp?punumber=6685222}
  {\emph {\bibinfo {title} {54th Annual {IEEE} Symposium on Foundations of
  Computer Science}}}\ (\bibinfo  {publisher} {{IEEE} Computer Society},\
  \bibinfo {year} {2013})\BibitemShut {NoStop}%
\bibitem [{\citenamefont {Strichman}\ and\ \citenamefont
  {Szeider}(2010)}]{DBLP:conf/sat/2010}%
  \BibitemOpen
  \bibinfo {editor} {\bibfnamefont {O.}~\bibnamefont {Strichman}}\ and\
  \bibinfo {editor} {\bibfnamefont {S.}~\bibnamefont {Szeider}},\ eds.,\ \href
  {\doibase 10.1007/978-3-642-14186-7} {\emph {\bibinfo {title} {Theory and
  Applications of Satisfiability Testing - {SAT} 2010, 13th International
  Conference, {SAT} 2010, Edinburgh, UK, July 11-14, 2010. Proceedings}}},\
  \bibinfo {series} {Lecture Notes in Computer Science}, Vol.\ \bibinfo
  {volume} {6175}\ (\bibinfo  {publisher} {Springer},\ \bibinfo {year}
  {2010})\BibitemShut {NoStop}%
\bibitem [{\citenamefont {Boufkhad}\ \emph {et~al.}(2005)\citenamefont
  {Boufkhad}, \citenamefont {Dubois}, \citenamefont {Interian},\ and\
  \citenamefont {Selman}}]{boufkhad05regular}%
  \BibitemOpen
  \bibfield  {author} {\bibinfo {author} {\bibfnamefont {Y.}~\bibnamefont
  {Boufkhad}}, \bibinfo {author} {\bibfnamefont {O.}~\bibnamefont {Dubois}},
  \bibinfo {author} {\bibfnamefont {Y.}~\bibnamefont {Interian}}, \ and\
  \bibinfo {author} {\bibfnamefont {B.}~\bibnamefont {Selman}},\ }\href
  {\doibase 10.1007/s10817-005-9012-z} {\bibfield  {journal} {\bibinfo
  {journal} {J. Autom. Reasoning}\ }\textbf {\bibinfo {volume} {35}},\ \bibinfo
  {pages} {181} (\bibinfo {year} {2005})}\BibitemShut {NoStop}%
\bibitem [{\citenamefont {Hoffmann}(2010{\natexlab{b}})}]{Hoffmann2010thesis}%
  \BibitemOpen
  \bibfield  {author} {\bibinfo {author} {\bibfnamefont {C.}~\bibnamefont
  {Hoffmann}},\ }\emph {\bibinfo {title} {Computational Complexity of Graph
  Polynomials}},\ \href@noop {} {Ph.D. thesis},\ \bibinfo  {school}
  {Universit\"{a}t des Saarlandes} (\bibinfo {year}
  {2010}{\natexlab{b}})\BibitemShut {NoStop}%
\bibitem [{\citenamefont {Averbouch}\ and\ \citenamefont
  {Makowsky}(2007)}]{averbouch2007complexity}%
  \BibitemOpen
  \bibfield  {author} {\bibinfo {author} {\bibfnamefont {I.}~\bibnamefont
  {Averbouch}}\ and\ \bibinfo {author} {\bibfnamefont {J.~A.}\ \bibnamefont
  {Makowsky}},\ }\href@noop {} {\bibfield  {journal} {\bibinfo  {journal}
  {preprint, January}\ } (\bibinfo {year} {2007})}\BibitemShut {NoStop}%
\bibitem [{\citenamefont {Alon}\ and\ \citenamefont
  {Spencer}(2008)}]{alon2008probabilistic}%
  \BibitemOpen
  \bibfield  {author} {\bibinfo {author} {\bibfnamefont {N.}~\bibnamefont
  {Alon}}\ and\ \bibinfo {author} {\bibfnamefont {J.~H.}\ \bibnamefont
  {Spencer}},\ }\href {\doibase 10.1002/9780470277331} {\emph {\bibinfo {title}
  {The probabilistic method}}},\ \bibinfo {edition} {3rd}\ ed.,\
  Wiley-Interscience Series in Discrete Mathematics and Optimization\ (\bibinfo
   {publisher} {John Wiley \& Sons, Inc., Hoboken, NJ},\ \bibinfo {year}
  {2008})\ pp.\ \bibinfo {pages} {xviii+352},\ \bibinfo {note} {with an
  appendix on the life and work of Paul Erd{\H{o}}s}\BibitemShut {NoStop}%
\bibitem [{\citenamefont {Erd{\H{o}}s}\ and\ \citenamefont
  {Lov{\'a}sz}(1975)}]{erdos1975problems}%
  \BibitemOpen
  \bibfield  {author} {\bibinfo {author} {\bibfnamefont {P.}~\bibnamefont
  {Erd{\H{o}}s}}\ and\ \bibinfo {author} {\bibfnamefont {L.}~\bibnamefont
  {Lov{\'a}sz}},\ }in\ \href@noop {} {\emph {\bibinfo {booktitle} {Infinite and
  finite sets ({C}olloq., {K}eszthely, 1973; dedicated to {P}. {E}rd{\H o}s on
  his 60th birthday), {V}ol. {II}}}}\ (\bibinfo  {publisher} {North-Holland,
  Amsterdam},\ \bibinfo {year} {1975})\ pp.\ \bibinfo {pages} {609--627.
  Colloq. Math. Soc. J\'anos Bolyai, Vol. 10}\BibitemShut {NoStop}%
\bibitem [{\citenamefont {Kratochv{\'{\i}}l}\ \emph {et~al.}(1993)\citenamefont
  {Kratochv{\'{\i}}l}, \citenamefont {Savick{\'{y}}},\ and\ \citenamefont
  {Tuza}}]{kratochvil93one}%
  \BibitemOpen
  \bibfield  {author} {\bibinfo {author} {\bibfnamefont {J.}~\bibnamefont
  {Kratochv{\'{\i}}l}}, \bibinfo {author} {\bibfnamefont {P.}~\bibnamefont
  {Savick{\'{y}}}}, \ and\ \bibinfo {author} {\bibfnamefont {Z.}~\bibnamefont
  {Tuza}},\ }\href {\doibase 10.1137/0222015} {\bibfield  {journal} {\bibinfo
  {journal} {{SIAM} J. Comput.}\ }\textbf {\bibinfo {volume} {22}},\ \bibinfo
  {pages} {203} (\bibinfo {year} {1993})}\BibitemShut {NoStop}%
\bibitem [{\citenamefont {Kitaev}\ \emph {et~al.}(2002)\citenamefont {Kitaev},
  \citenamefont {Shen},\ and\ \citenamefont {Vyalyi}}]{kitaev2002classical}%
  \BibitemOpen
  \bibfield  {author} {\bibinfo {author} {\bibfnamefont {A.~Y.}\ \bibnamefont
  {Kitaev}}, \bibinfo {author} {\bibfnamefont {A.~H.}\ \bibnamefont {Shen}}, \
  and\ \bibinfo {author} {\bibfnamefont {M.~N.}\ \bibnamefont {Vyalyi}},\
  }\href {\doibase 10.1090/gsm/047} {\emph {\bibinfo {title} {Classical and
  quantum computation}}},\ \bibinfo {series} {Graduate Studies in Mathematics},
  Vol.~\bibinfo {volume} {47}\ (\bibinfo  {publisher} {American Mathematical
  Society, Providence, RI},\ \bibinfo {year} {2002})\ pp.\ \bibinfo {pages}
  {xiv+257},\ \bibinfo {note} {translated from the 1999 Russian original by
  Lester J. Senechal}\BibitemShut {NoStop}%
\bibitem [{\citenamefont {Levit}\ and\ \citenamefont
  {Mandrescu}(2005)}]{levit2005independence}%
  \BibitemOpen
  \bibfield  {author} {\bibinfo {author} {\bibfnamefont {V.~E.}\ \bibnamefont
  {Levit}}\ and\ \bibinfo {author} {\bibfnamefont {E.}~\bibnamefont
  {Mandrescu}},\ }in\ \href {http://cai05.web.auth.gr/papers/20.pdf} {\emph
  {\bibinfo {booktitle} {Proceedings of the 1st {I}nternational {C}onference on
  {A}lgebraic {I}nformatics}}}\ (\bibinfo  {publisher} {Aristotle Univ.
  Thessaloniki, Thessaloniki},\ \bibinfo {year} {2005})\ pp.\ \bibinfo {pages}
  {231--254}\BibitemShut {NoStop}%
\bibitem [{\citenamefont {Aharonov}\ and\ \citenamefont
  {Naveh}(2002)}]{quant-ph/0210077}%
  \BibitemOpen
  \bibfield  {author} {\bibinfo {author} {\bibfnamefont {D.}~\bibnamefont
  {Aharonov}}\ and\ \bibinfo {author} {\bibfnamefont {T.}~\bibnamefont
  {Naveh}},\ }\href@noop {} {\enquote {\bibinfo {title} {Quantum np - a
  survey},}\ } (\bibinfo {year} {2002}),\ \Eprint
  {http://arxiv.org/abs/arXiv:quant-ph/0210077} {arXiv:quant-ph/0210077}
  \BibitemShut {NoStop}%
\bibitem [{\citenamefont {Cook}(1971)}]{Cook:1971:CTP:800157.805047}%
  \BibitemOpen
  \bibfield  {author} {\bibinfo {author} {\bibfnamefont {S.~A.}\ \bibnamefont
  {Cook}},\ }in\ \href {\doibase 10.1145/800157.805047} {\emph {\bibinfo
  {booktitle} {Proceedings of the Third Annual ACM Symposium on Theory of
  Computing}}},\ \bibinfo {series and number} {STOC '71}\ (\bibinfo
  {publisher} {ACM},\ \bibinfo {address} {New York, NY, USA},\ \bibinfo {year}
  {1971})\ pp.\ \bibinfo {pages} {151--158}\BibitemShut {NoStop}%
\bibitem [{\citenamefont {Dhar}(1983)}]{PhysRevLett.51.853}%
  \BibitemOpen
  \bibfield  {author} {\bibinfo {author} {\bibfnamefont {D.}~\bibnamefont
  {Dhar}},\ }\href {\doibase 10.1103/PhysRevLett.51.853} {\bibfield  {journal}
  {\bibinfo  {journal} {Phys. Rev. Lett.}\ }\textbf {\bibinfo {volume} {51}},\
  \bibinfo {pages} {853} (\bibinfo {year} {1983})}\BibitemShut {NoStop}%
\bibitem [{\citenamefont {Hsu}\ \emph {et~al.}(2013)\citenamefont {Hsu},
  \citenamefont {Laumann}, \citenamefont {L\"auchli}, \citenamefont
  {Moessner},\ and\ \citenamefont {Sondhi}}]{PhysRevA.87.062334}%
  \BibitemOpen
  \bibfield  {author} {\bibinfo {author} {\bibfnamefont {B.}~\bibnamefont
  {Hsu}}, \bibinfo {author} {\bibfnamefont {C.~R.}\ \bibnamefont {Laumann}},
  \bibinfo {author} {\bibfnamefont {A.~M.}\ \bibnamefont {L\"auchli}}, \bibinfo
  {author} {\bibfnamefont {R.}~\bibnamefont {Moessner}}, \ and\ \bibinfo
  {author} {\bibfnamefont {S.~L.}\ \bibnamefont {Sondhi}},\ }\href {\doibase
  10.1103/PhysRevA.87.062334} {\bibfield  {journal} {\bibinfo  {journal} {Phys.
  Rev. A}\ }\textbf {\bibinfo {volume} {87}},\ \bibinfo {pages} {062334}
  (\bibinfo {year} {2013})}\BibitemShut {NoStop}%
\bibitem [{\citenamefont {Itzykson}\ \emph {et~al.}(1986)\citenamefont
  {Itzykson}, \citenamefont {Saleur},\ and\ \citenamefont
  {Zuber}}]{Itzykson1986}%
  \BibitemOpen
  \bibfield  {author} {\bibinfo {author} {\bibfnamefont {C.}~\bibnamefont
  {Itzykson}}, \bibinfo {author} {\bibfnamefont {H.}~\bibnamefont {Saleur}}, \
  and\ \bibinfo {author} {\bibfnamefont {J.-B.}\ \bibnamefont {Zuber}},\ }\href
  {http://stacks.iop.org/0295-5075/2/i=2/a=004} {\bibfield  {journal} {\bibinfo
   {journal} {EPL (Europhysics Letters)}\ }\textbf {\bibinfo {volume} {2}},\
  \bibinfo {pages} {91} (\bibinfo {year} {1986})}\BibitemShut {NoStop}%
\bibitem [{\citenamefont {Kirkpatrick}\ and\ \citenamefont
  {Selman}(1994)}]{Kirkpatrick27051994}%
  \BibitemOpen
  \bibfield  {author} {\bibinfo {author} {\bibfnamefont {S.}~\bibnamefont
  {Kirkpatrick}}\ and\ \bibinfo {author} {\bibfnamefont {B.}~\bibnamefont
  {Selman}},\ }\href {\doibase 10.1126/science.264.5163.1297} {\bibfield
  {journal} {\bibinfo  {journal} {Science}\ }\textbf {\bibinfo {volume}
  {264}},\ \bibinfo {pages} {1297} (\bibinfo {year} {1994})}\BibitemShut
  {NoStop}%
\bibitem [{\citenamefont {Kurtze}\ and\ \citenamefont
  {Fisher}(1979)}]{PhysRevB.20.2785}%
  \BibitemOpen
  \bibfield  {author} {\bibinfo {author} {\bibfnamefont {D.~A.}\ \bibnamefont
  {Kurtze}}\ and\ \bibinfo {author} {\bibfnamefont {M.~E.}\ \bibnamefont
  {Fisher}},\ }\href {\doibase 10.1103/PhysRevB.20.2785} {\bibfield  {journal}
  {\bibinfo  {journal} {Phys. Rev. B}\ }\textbf {\bibinfo {volume} {20}},\
  \bibinfo {pages} {2785} (\bibinfo {year} {1979})}\BibitemShut {NoStop}%
\bibitem [{\citenamefont {Levin}(1973)}]{zbMATH03489106}%
  \BibitemOpen
  \bibfield  {author} {\bibinfo {author} {\bibfnamefont {L.}~\bibnamefont
  {Levin}},\ }\href@noop {} {\bibfield  {journal} {\bibinfo  {journal} {Probl.
  Peredachi Inf.}\ }\textbf {\bibinfo {volume} {9}},\ \bibinfo {pages} {115}
  (\bibinfo {year} {1973})}\BibitemShut {NoStop}%
\bibitem [{\citenamefont {Mezard}\ \emph {et~al.}(2002)\citenamefont {Mezard},
  \citenamefont {Parisi},\ and\ \citenamefont {Zecchina}}]{Mezard02082002}%
  \BibitemOpen
  \bibfield  {author} {\bibinfo {author} {\bibfnamefont {M.}~\bibnamefont
  {Mezard}}, \bibinfo {author} {\bibfnamefont {G.}~\bibnamefont {Parisi}}, \
  and\ \bibinfo {author} {\bibfnamefont {R.}~\bibnamefont {Zecchina}},\ }\href
  {\doibase 10.1126/science.1073287} {\bibfield  {journal} {\bibinfo  {journal}
  {Science}\ }\textbf {\bibinfo {volume} {297}},\ \bibinfo {pages} {812}
  (\bibinfo {year} {2002})}\BibitemShut {NoStop}%
\bibitem [{\citenamefont {Monasson}\ \emph {et~al.}(1999)\citenamefont
  {Monasson}, \citenamefont {Zecchina}, \citenamefont {Kirkpatrick},
  \citenamefont {Selman},\ and\ \citenamefont {Troyansky}}]{Monasson1999}%
  \BibitemOpen
  \bibfield  {author} {\bibinfo {author} {\bibfnamefont {R.}~\bibnamefont
  {Monasson}}, \bibinfo {author} {\bibfnamefont {R.}~\bibnamefont {Zecchina}},
  \bibinfo {author} {\bibfnamefont {S.}~\bibnamefont {Kirkpatrick}}, \bibinfo
  {author} {\bibfnamefont {B.}~\bibnamefont {Selman}}, \ and\ \bibinfo {author}
  {\bibfnamefont {L.}~\bibnamefont {Troyansky}},\ }\href
  {http://dx.doi.org/10.1038/22055} {\bibfield  {journal} {\bibinfo  {journal}
  {Nature}\ }\textbf {\bibinfo {volume} {400}},\ \bibinfo {pages} {133}
  (\bibinfo {year} {1999})}\BibitemShut {NoStop}%
\bibitem [{\citenamefont {Parisi}\ and\ \citenamefont
  {Sourlas}(1981)}]{PhysRevLett.46.871}%
  \BibitemOpen
  \bibfield  {author} {\bibinfo {author} {\bibfnamefont {G.}~\bibnamefont
  {Parisi}}\ and\ \bibinfo {author} {\bibfnamefont {N.}~\bibnamefont
  {Sourlas}},\ }\href {\doibase 10.1103/PhysRevLett.46.871} {\bibfield
  {journal} {\bibinfo  {journal} {Phys. Rev. Lett.}\ }\textbf {\bibinfo
  {volume} {46}},\ \bibinfo {pages} {871} (\bibinfo {year} {1981})}\BibitemShut
  {NoStop}%
\bibitem [{\citenamefont {Scott}\ and\ \citenamefont
  {Sokal}(2006)}]{Scott:2006:DGL:1108656.1108666}%
  \BibitemOpen
  \bibfield  {author} {\bibinfo {author} {\bibfnamefont {A.~D.}\ \bibnamefont
  {Scott}}\ and\ \bibinfo {author} {\bibfnamefont {A.~D.}\ \bibnamefont
  {Sokal}},\ }\href {\doibase 10.1017/S0963548305007182} {\bibfield  {journal}
  {\bibinfo  {journal} {Comb. Probab. Comput.}\ }\textbf {\bibinfo {volume}
  {15}},\ \bibinfo {pages} {253} (\bibinfo {year} {2006})}\BibitemShut
  {NoStop}%
\bibitem [{\citenamefont {Todo}(1997)}]{cond-mat/9703176}%
  \BibitemOpen
  \bibfield  {author} {\bibinfo {author} {\bibfnamefont {S.}~\bibnamefont
  {Todo}},\ }\href@noop {} {\enquote {\bibinfo {title} {Transfer-matrix study
  of hard-core singularity for hard-square lattice gas},}\ } (\bibinfo {year}
  {1997}),\ \Eprint {http://arxiv.org/abs/arXiv:cond-mat/9703176}
  {arXiv:cond-mat/9703176} \BibitemShut {NoStop}%
\end{thebibliography}%
\nocite{*}
\appendix
\section{Proof of Theorem~\ref{thm:qshearer_qsat}}
\label{sec:shearer_proof}
\begin{definition} For a subspace $X \subset V$, let $\RR(X)=\frac{\dim(X)}{\dim(V)}$. We say that $X$ is mutually R-independent of the subspaces $Y_1,\ldots,Y_n \subset V$ if $\forall S \subset [n],\  \RR\left( X \bigcap_{i \in S} Y_i \right)= \RR(X) \RR\left( \bigcap_{i \in S} Y_i \right)$.
\end{definition}

\begin{definition}
The subspaces $X_{1},\ldots,X_{n}$ have R-dependency graph $G=([n],E)$ if $X_{i}$ is mutually R-independent of  $\{X_j \}_{ j\in [n] \setminus \Gamma^+(i) }$ where $\Gamma(i) \equiv \{ j | (i,j) \in E \}$,  $\Gamma^+(i) \equiv \Gamma(i) \cup \{i\}$. 
\end{definition}

An independent set (also known as a stable set) of a graph $G$ is a set of non-adjacent vertices. $\textnormal{Indep}(G)$ is the family of all the independent sets in $G$.
The independent set polynomial is
\begin{equation*}
 I(G,x) =   \sum_{S \in \text{Indep}(G)} x^{|S|} 
\end{equation*}
We use the shorthand $I(G)=I(G,-p)$ whenever $p$ is clear from the context.

\begin{lemma}[\cite{Scott2005}] Fix $(G,p)$. The following properties are equivalent:
\begin{enumerate}
\item For every $0 \leq p'\leq p,\ I(G,-p')> 0 $.
\item For every induced subgraph $F$ of $G,\  I(F,-p)> 0$.
\end{enumerate}
If this is the case, we say that $(G,p)$ has Shearer's property.
\label{le:shearer_property}
\end{lemma}

\begin{theorem}

Let $X_{1},\ldots ,X_{n}$ subspaces of $V$ with dependency graph $G$, such that $\forall i \in [n],\ \RR(X_{i})\geq 1- p$. If $(G,p)$ has Shearer's property, then $\RR(\bigcap_{i \in [n]} X_i) \geq I(G,p) > 0 $.
\label{thm:qshearer}
\end{theorem}
For comparison, the following is the classical analogue by Shearer~\cite{Shearer1985}.
\begin{theorem}
Let $B_1,\ldots,B_n$ be events with dependency graph $G$,
such that $\forall i \in [n]\,\ \Pr(B_i) \geq 1-p$.
 If $(G,p)$ has Shearer's property, then $\Pr(\bigcap_{i \in [n]} B_i) \geq I(G,p) > 0 $.
\end{theorem}
\begin{proofof}{Theorem~\ref{thm:qshearer}}
We will first introduce some notations. For $S \subset [n]$, we use the shorthand $X_S=\bigcap_{i \in S} X_i$, where $X_\emptyset = V$. The subgraph induced by $U \subset V(G)$ is denoted $G[U]$.
\begin{lemma} For any $S \subset T \subset [n]$,
\[  \frac{\RR(X_S)}{I(G[S])} \leq  \frac{\RR(X_T)}{I(G[T])}  \]
\end{lemma}
The lemma completes the proof of the theorem, using $S=\emptyset$ and $T=[n]$: $\frac{\RR(X_{[n]}) }{I(G) }\geq \frac{\RR(X_\emptyset)}{I(G[\emptyset])}=1$.

We prove the lemma by induction on the size of $S$ and assume that $T=S \cup \{ i \}$.
Let $A= S \setminus \Gamma(i)$. 
By partitioning the independent sets in $T$ to ones that contain $i$ and the ones that do not, we have
\begin{equation}
I(G[T])=I(G[S])-p I(G[A]).
\label{eq:I_GT}
\end{equation}

We get a lower bound similar to the previous equation

\begin{align}
\RR(X_{T})&=\RR(X_{\Gamma(i)} \cap X_{i} \cap X_{A} )\nonumber \\
       & \geq \RR(X_{\Gamma(i)} \cap X_{A}) + \RR( X_i \cap X_{A} ) - \RR(X_{A}) \nonumber \\
			 & = \RR(X_{S}) + \RR(X_{i})\RR(X_{A})-\RR(X_{A}) \nonumber\\
			 & \geq \RR(X_{S}) - p\RR(X_{A})
\label{eq:RX_T}			 
\end{align}

The first inequality follows from Fact~\ref{fa:xcapycapz}, the second equality follows from the fact that $X_{i}$ is R-independent of $\{X_{j}\}_{j\in A}$, and the last inequality follows from the assumption $\RR(X_{i}) \geq 1 - p$.

\begin{fact}
\[\RR(X \cap Y \cap Z) \geq \RR(X \cap Z) + \RR(Y \cap Z) - \RR(Z) .\]
\label{fa:xcapycapz}
\end{fact}
Fact~\ref{fa:xcapycapz} follows from the following property by assigning $A=X \cap Z, B=Y \cap Z$:
\[ \dim(A \cap B) = \dim(A)+\dim(B) - \dim(A + B) \] for any two subspaces $A,B \subset V$, where $A+B = \{a+b|a\in A, b\in B\} $. 

To complete the proof of the lemma,
\begin{align*}
\frac{\RR(X_{T})}{I(G[T])}& -\frac{\RR(X_{S})}{I(G[S])} \geq  \frac{\RR(X_{S}) - p \RR(X_{A})}{I(G[S])-p I(G[A])}-\frac{\RR(X_{S})}{I(G[S])} \\
 & =\frac{p(I(G[A])\RR(X_{S}) - \RR(X_{A})I(G[S]) )} {I(G[S])(I(G[S]) - p I(G[A]))} \\
 &= \frac{p I(G[A])}{I(G[S])-p I(G[A])} \left(\frac{\RR(X_{S})}{I(G[S])} - \frac{\RR(X_I)}{I(G[A])} \right) \\
 & =  \frac{p I(G[A])}{I(G[T])} \left(\frac{\RR(X_{S})}{I(G[S])} - \frac{\RR(X_I)}{I(G[A])} \right)  \geq 0 ,
\end{align*}
where we used Eq.~\eqref{eq:I_GT} and ~\eqref{eq:RX_T}  in the first inequality, Eq.~\eqref{eq:I_GT} in the third equality, and the second definition of Shearer's property (see Lemma~\ref{le:shearer_property}) and the induction hypothesis in the last inequality.

\end{proofof}

\begin{proofof}{Theorem~\ref{thm:qshearer_qsat}}
This is an application of Theorem~\ref{thm:qshearer}. For a QSAT instance $\Pi_1,\ldots,\Pi_m$, with interaction graph $F$ we define $X_i = \ker(\Pi_i)$, and $G$ has $m$ vertices, and an edge between $i$ and $j$ if $\Pi_i$ and $\Pi_j$ act on a shared qudit. $G$ is indeed an $R$-dependency graph for $X_1,\ldots,X_m$: this follows from the property $\dim(A \otimes B)=\dim(A) \dim(B)$~\cite{AmbainisKS12}.

To complete the proof we observe that $I(G,-p)=\ZZ(F,-p)$ (in graph theoretic terminology, this is equivalent to the statement that for any hypergraph $F$, $I_{L(F)}(x)= M_F(x)$,  where $L(F)$ is the line graph of $F$ and $M_F(x)$ is the matching polynomial of $F$, see ~\cite{levit2005independence}).
\end{proofof}

\section{The cavity derivations}
\label{sec:cavity_derivations}

The cavity method is a collection of techniques for calculating the  properties of statistical mechanical models defined on locally tree-like lattices, \emph{i.e.} on graphs without short loops. Here, we apply it to evaluating the free energy $F = \log \ZZ$ of the hard-core gas of particles living on the hyperedges $i$ of the interaction graph $G$, as defined by Eq.~\eqref{eq:Zhardcore}.

We introduce a pair of cavity distributions (`messages' or `beliefs'), $p_{i \to a}(n_i)$ and $p_{a\to i}(n_i)$, for each link between a hyperedge $i$ and node $a$. These represent the marginal distribution of the occupation variable $n_i$ in the absence of its link to site $a$ and in the absence of its link to sites other than $a$, respectively. On a graph with no loops, these distributions are simply the normalized partial sums of $\mathcal{Z}$ when evaluated from the leaves inward towards the link between $a$ and $i$.

Assuming that the cavity distributions coming into a given node or edge are independent (as they are on graphs with no loops), the $2 M k$ cavity messages satisfy the local relations,
\begin{align*}
p_{a\rightarrow i}^{'}\left(n_{i}\right) & =\sum_{\substack{\{n_{j}\},\\
j\epsilon\partial a\backslash i
}
}\left[\mathbb{I}\left(\sum_{k\epsilon\partial a}n_{k}\leq1\right)\right]\lambda^{n_{i}}\prod_{j\epsilon\partial a\backslash i}p_{j\rightarrow a}\left(n_{j}\right)\\
p_{i\rightarrow a}^{'}\left(n_{i}\right) & =\lambda^{n_{i}}\prod_{b\epsilon\partial i\backslash a}\left(\dfrac{p_{b\rightarrow i}\left(n_{i}\right)}{\lambda^{n_{i}}}\right)
\end{align*}
where $p'$ denotes an unnormalized distribution. Properly normalized, each cavity distribution can be parametrized by one real number, which we take to be the probability of occupation (recall that $n_i \in \{0,1\}$),
\begin{align*}
q_{a\rightarrow i} &= p_{a\to i}(1) &
l_{i\rightarrow a} &= p_{i\to a}(1)
\end{align*}
Using this parametrization leads directly to the BP equations in the main text.

On graphs without loops, the BP equations admit a single global solution for the cavity messages $q_{a\rightarrow i}$ and $l_{i\rightarrow a}$, which corresponds to an exact evaluation of $\ZZ$. From this solution, one obtains the total free energy Eq.~\eqref{eq:free_energies} as a sum of local contributions due to the free energy of each hyperedge $F_i$, site $F_a$ and link $F_{ai}$. These free energies are functions of the messages coming into the relevant piece of the interaction graph. Explicitly,
\begin{align*}
F_{a} & ={\displaystyle \log\left[\sum_{\substack{\{n_{j}\},\\
j\epsilon\partial a
}
}\prod_{j}p_{j\rightarrow a}\left(n_{j}\right)\right]}\notag\\
 & =\log\left[{\displaystyle \prod_{j\epsilon\partial a}\left(1-l_{j\rightarrow a}\right)+\sum_{j\epsilon\partial a}l_{j\rightarrow a}\prod_{k\epsilon\partial a\backslash j}\left(1-l_{k\rightarrow a}\right)}\right]\\
F_{i} & =\log\left[{\displaystyle \sum_{n_{i}}\lambda^{n_{i}}\prod_{b\epsilon\partial i}\dfrac{p_{b\rightarrow i}\left(n_{i}\right)}{\lambda^{n_{i}}}}\right]\notag\\
 & =\log\left[{\displaystyle \prod_{b\epsilon\partial i}\left(1-q_{b\rightarrow i}\right)+\lambda\prod_{b\epsilon\partial i}\dfrac{q_{b\rightarrow i}}{\lambda}}\right]\\
F_{ai} & =\log\left[{\displaystyle \sum_{n_{i}}\dfrac{1}{\lambda^{n_{i}}}p_{i\rightarrow a}\left(n_{i}\right)p_{a\rightarrow i}\left(n_{i}\right)}\right]\notag\\
 & =\log\left[{\displaystyle \left(1-l_{i\rightarrow a}\right)\left(1-q_{a\rightarrow i}\right)+\dfrac{l_{i\rightarrow a}q_{a\rightarrow i}}{\lambda}}\right]
\end{align*}
This evaluation of the free energy is exact for finite trees. 
For infinite graphs with loops of divergent girth, it often produces asymptotically exact results.

Thus, the main task in evaluating the free energy (and corresponding $\ZZ$) for a given infinite graph is to obtain solutions of the infinite set of BP equations for the cavity messages. 

\subsection{Solving the cavity equations: regular graphs}

For infinite $t$-regular $k$-local trees/regular random graphs, the BP equations admit a solution with uniform messages $q_{a\to i} = q$ and $l_{i \to a} = l$. After a small amount of algebra, the self-consistent equations can be reduced to finding the roots of
\begin{align}
\label{eq:regroots}
x^k - x^{k-1} - \lambda(t-1) &= 0
\end{align}
where
\begin{align*}
x &= 1 + (t-1) \left( \dfrac{1}{l^{-1}-1} \right) \\
&= \lambda(q^{-1} - 1)
\end{align*}
The discriminant of Eq.~\eqref{eq:regroots} is (with $z = \lambda(t-1)$),

 \begin{align*}
\Delta &= (-1)^{\binom{k}{2}}\left| 
  \left(
   \begin{BMAT}{cccccc.ccccc}{cccc.cccc}
  1 & -1 & 0 & \cdots & 0&  -z &                 0& &\cdots & & 0 \\
  0 & 1 & -1          & 0 & \cdots & 0&          -z & 0           & \cdots & & 0 \\
     &      & \vdots&     &    &     &                     & & \vdots            & &  \\
   &0  &  \cdots &1 &-1 & 0 &                        0 & & \cdots & & -z  \\
  k & 1-k & 0 & \cdots & 0&  0 &0& &\cdots & & 0 \\
  0 & k & 1-k & 0 & \cdots & 0  & 0 &  & \cdots & & 0 \\
  & & \vdots&  &  & && &\vdots & & \\
  &0  &  \cdots & &k &1- k  &0 & & \cdots & & 0  
 \end{BMAT}
 \right) \right|  \nonumber \\[-1ex]
& \qquad \qquad \qquad      \hexbrace{4.5cm}{k+1} \ \   \hexbrace{2.1cm}{k-2} \\
& =  (-1)^{\binom{k}{2}}(-z)^{k-2} \left[-z k^k  + (-1)^k \left( (1-k)^k +k (1-k)^{k-1} \right) \right] \nonumber \\
 & = (-1)^{\binom{k}{2} + k + 1} z^{k-2}((k-1)^{k-1}+k^k z)
 \end{align*}
where in the second step we observed that only 3 terms in the determinant are non-zero. The discriminant has a simple zero at $z_c = -\frac{(k-1)^{k-1}}{k^k}$, in agreement with Eq.~\eqref{eq:lambdacreg}. 
The zeros of the discriminant indicate that values of $z$ where two roots of the BP equations merge so that there is a non-analyticity in the resulting free energy. 

At $z_c$, the root of interest has multiplicity 2 and can thus be easily found by looking for a root shared with the derivative of Eq.~\eqref{eq:regroots}. 
Thus, 
\begin{align*}
x_c = \frac{k-1}{k}
\end{align*} 
which can be verified by direct evaluation. Geometrically, we expect $x$ to evolve as a square root of deviations in $z$ from criticality. To leading non-vanishing order in $\delta z = z - z_c$ and $\delta x = x - x_c$,  Eq.~\eqref{eq:regroots} can be solved,
\begin{align}
\label{eq:devroot}
    \delta z&= \frac{1}{2} (\delta x)^2 \left[ \left(\frac{k-1}{k}\right)^{k-3}(k-1) \right]
\end{align}

From the critical values for $z_c$ and $x_c$, and the behavior near criticality Eq.~\eqref{eq:devroot}, it is a straightforward if algebraically tedious process to evaluate the free energy density given in Eq.~\eqref{eq:freeenergytree}.


\subsection{Solving the cavity equations: random graphs}

In infinite Erd\H{o}s-R\'enyi random graphs, the local geometry fluctuates from site to site and, accordingly, the local BP equations vary. 
We do not expect to find a solution with uniform messages; 
rather, the \emph{distribution} of cavity messages $P_q[\cdot]$ and $P_l[\cdot]$ ought to be stable under the iteration of the BP equations. 
This leads to self-consistent equations for the distributions $P_q[\cdot]$ and $P_l[\cdot]$, which we solve by numerical population dynamics. 

\section{Local degree fluctuations for random \texorpdfstring{$k$}{k}-QSAT:}
\label{sec:degree_fluctuations_for_random_qsat}
In an Erd\H{o}s-R\'enyi random graph, the degree $z$ of each site follows a Poisson distribution, i.e.,
\begin{align*}
\Pr(\mathrm{degree}=z)=\mathrm{e}^{-k\alpha}\dfrac{\left(k\alpha\right)^{z}}{z!} \ .
\end{align*}
This means that there is a small but finite density of large degree sites at any $\alpha$. 
These large degree sites pose a problem for the Shearer bound: 
the partition function of the finite `star' of hyperedges surrounding a site of degree $z$ is $\ZZ = 1 + z \lambda$. This has its first negative fugacity zero at $\lambda_c = -1/z$. 
Given the unbounded fluctuations in $z$, monotonicity implies that $\lambda_c = 0$ 
for the infinite graph. 

However, we believe that it is not these fluctuations which in fact
lead to the instance becoming UNSAT for the case of qubits. This can already be seen classically:
one can explicitly verify that a star which by Shearer should be UNSAT will, in fact, always be
SAT; this happens because the constraints imposed by different projectors on the central spin,
can never be disjoint as the worst-case Shearer bounds assumes. 

This carries over to the quantum case. Indeed, here it has been shown \cite{BravyiMR10} that
stars are not only satisfiable, but even have entropic ground spaces $\dim \ker H = 2(2^{k-1}-1)^z (\frac{z}{2^k-2} + 1)$.  
This is backed up by the rigorous work of \cite{AmbainisKS12}, who find a lower bound on the satisfiability threshold by  gluing together product states on high degree regions of the Erd\H{o}s-R\'enyi graph with states guaranteed by the QLLL on the low degree regions.

Our numerics, in turn, also implicitly neglect the large degree fluctuations because the density of sites with $z > 2^k$ is very small ($10^{-14}$ at $k=5$ to $10^{-370}$ at $k=9$) for $\alpha \approx \alpha_c$ of the observed bulk transition. 
By contrast, for $k < 5$, the large degree fluctuations are not as rare and thus the population dynamics fails to converge.

For $k \ge 5$, the population dynamics converges up to some $\lambda_c(\alpha)$ and, moreover, we clearly observe
 the square root singularity in $\langle n \rangle$ referenced to this $\lambda_c$. This suggests that we are detecting a bulk transition in the lattice gas, from which we determine $\alpha_c$ by solving $\lambda_c(\alpha) = 1/2^k$, see Fig.~\ref{fig:randomQSAT_transition}. 

\section{ Improved bounds using the critical exponents}
\label{sec:critical_exponents}
We can rewrite $R(\ker H) \geq \exp(nf)=p_{eff}^n$, where $p_{eff}=\exp(f)$.
In the large $n$ limit, there exists a (non-analytic) power series expansion of $f(\lambda)$ near $\lambda_c$. 
We learn a few interesting facts from the existence of this power series expansion (and in particular, that $f$ does not diverge at the critical point). First, $p_{eff}\neq 0$ at the critical point. Second, $f$ is known to have non-analytic terms, that depend only on the lattice dimension. For example, for all 1-D lattices (chains, ladders, etc.), the expansion close to the transition point is  $f(\lambda)= a_0 + a_{1/2} (\lambda - \lambda_c)^{1/2}+ \mathcal{O}(\lambda - \lambda_c)$, and for 2-D lattices, $f(\lambda)=b_0 + b_{5/6} (\lambda-\lambda_c)^{5/6} + \mathcal{O}(\lambda-\lambda_c)$.

To conclude, near the critical point, the leading power of the non-analytic part of $f$ depend only on the dimension of the lattice involved, and not on the specific details of the lattice. As a result, the dimension of the satisfying subspace grows \emph{faster} than one would expect for 1-D and 2-D lattices.  

\end{document}